\documentclass[prd,11pt,floatfix,superscriptaddress]{revtex4}

\usepackage{latexsym,amsbsy}
 \usepackage[dvips]{graphicx}
 \DeclareGraphicsExtensions{.eps, .jpg}

\begin{document}
\title{\boldmath
OBSERVATIONAL CONSTRAINTS\\ ON COSMOLOGICAL MODELS WITH CHAPLYGIN
GAS\\ AND QUADRATIC EQUATION OF STATE }



\author{G. S. Sharov} 

\affiliation{Tver state university\\ 170002, Sadovyj per. 35,
Tver, Russia}

\email{german.sharov@mail.ru} 

\begin{abstract}
Observational manifestations of accelerated expansion of the universe, in particular,
recent data for Type Ia supernovae, baryon acoustic oscillations,  for the Hubble
parameter $H(z)$ and cosmic microwave background constraints are described with
different cosmological models. We compare
 the $\Lambda$CDM, the models with generalized and modified Chaplygin gas
and the model  with quadratic equation of state. For these models we estimate optimal
model parameters and their permissible errors with different approaches to calculation
of sound horizon scale $r_s(z_d)$. Among the considered models the best value of
$\chi^2$ is achieved for the model with quadratic equation of state, but it has 2
additional parameters in comparison with the $\Lambda$CDM and therefore is not favored
by the Akaike information criterion.
\end{abstract}

\keywords{observational data; cosmological models; Chaplygin gas;
quadratic equation of state}

 \maketitle \flushbottom

\section{Introduction}\label{Intr}

Observations  \cite{Riess98,Perl99} of Type Ia supernovae demonstrated accelerated
expansion of our universe. Further investigations of supernovae
\cite{SNTable,WeinbergAcc12},  baryon acoustic oscillations
\cite{WeinbergAcc12,Eisen05,EisenHu98,Aubourg14}, cosmic microwave background
measurements \cite{Aubourg14,WMAP,Plank13,Plank15}, estimations
 \cite{Simon05,Stern10,Moresco12,Zhang12,Moresco15,Gazta09,Blake12,Busca12,ChuangW12,Chuang13,Anderson13,Anderson14,Oka13,Font-Ribera13,Delubac14}
 of the Hubble parameter
$H(z)$ for different redshifts $z$   confirmed accelerated growth of the cosmological
scale factor $a(t)$ at late stage of its evolution.

For Type Ia supernovae we can measure their redshifts $z$ and luminosity distances
$D_L$, so these objects may be used as standard candles
\cite{Riess98,Perl99,SNTable,WeinbergAcc12}.

Baryon acoustic oscillations (BAO) are  observed as a peak in the correlation function
of the galaxy distribution at the comoving sound horizon scale $r_s(z_d)$
\cite{Eisen05,EisenHu98}, corresponding to the end of the drag era, when baryons became
decoupled and acoustic waves propagation was ended. This effect has various
observational
 manifestations  \cite{WMAP,Plank13,Plank15,Gazta09,Blake12,Busca12,ChuangW12,Chuang13,Anderson13,Anderson14,Oka13,Font-Ribera13,Delubac14,Percival09,Kazin09,Beutler11,BlakeBAO11,Padmanabhan12,Seo12,Kazin14,Ross14},
in particular, one can estimate the Hubble parameter $H(z)$ for definite redshifts
 \cite{Gazta09,Blake12,Busca12,ChuangW12,Chuang13,Anderson13,Anderson14,Oka13,Font-Ribera13,Delubac14}
(details are in Sect.~\ref{Observ}).

The mentioned recent observations of Type Ia supernovae, BAO  and $H(z)$ essentially
restrict possible cosmological theories and models. To satisfy these observations all
models are to describe accelerated expansion of the universe with definite parameters
\cite{WMAP,Plank13,Plank15,CopelandST06,Clifton,Bamba12}.

The standard Einstein gravity with $\Lambda=0$ predicts deceleration of the expanding
universe: $a''(t)<0$. So to explain observed accelerated expansion, we are to modify
this theory. The most simple (and most popular) modification is the $\Lambda$CDM,
including dark energy  corresponding to $\Lambda\ne0$ and cold dark matter in addition
to deficient visible matter. This model with appropriate parameters
\cite{WeinbergAcc12,WMAP,Plank13,Plank15} successfully describes practically all
observational data, in Sect.~\ref{LCDM} we apply this model to describe the updated
recent observations of Type Ia supernovae, BAO effects and $H(z)$ estimates. In this
paper we use the notation $\Lambda$CDM for the model with an arbitrary spatial curvature
$\Omega_k$.

One should note that there are some problems in the $\Lambda$CDM model, in particular,
ambiguous nature of dark matter and dark energy, the problem of fine tuning for the
observed value of $\Lambda$ and the coincidence problem for surprising proximity
$\Omega_\Lambda$ and $\Omega_m$ nowadays  \cite{Clifton,Bamba12}.

Therefore cosmologists suggested a lot of alternative models with different equations of
state, scalar fields, with $f(R)$ Lagrangian, additional space dimensions and many
others
 \cite{Clifton,Bamba12,NojOdinFR,GrSh13,ShV14}. In this paper we
consider in detail two models with nontrivial equations of state describing both dark
matter and dark energy: the model with modified Chaplygin gas (MCG)
\cite{KamenMP01,Benaoum02,Chimento04,DebnathBCh04,LiuLiC05,LuXuWL11,PaulTh13,PaulTh14}
in Sect.~\ref{Chap} and the model  with quadratic equation of state
\cite{NojOdin04,NojOdinTs05,AnandaB06,LinderS09,AstashNojOdinY12,Chavanis13} in
Sect.~\ref{Quad}.

\section{Observational data}\label{Observ}

In this paper we use  the Union2.1 compilation \cite{SNTable} of Type Ia supernovae (SNe
Ia) observational data. This table includes redshifts $z=z_i$ and distance moduli
$\mu_i=\mu_i^{obs}$ with errors $\sigma_i$ for $N_{SN}=580$ supernovae. The distance
modulus $\mu_i=\mu(D_L)=5\log_{10}\big(D_L/10\mbox{pc}\big)$ is logarithm of the
luminosity distance \cite{Riess98,CopelandST06,Clifton}:
 \begin{equation}
 D_L(z)=\frac{c\,(1+z)}{H_0}S_k
 \bigg(H_0\int\limits_0^z\frac{d\tilde z}{H(\tilde
 z)}\bigg).  \label{DL} \end{equation}
 Here

\vspace{-5mm}

$$S_k(x)=\left\{\begin{array}{ll} \sinh\big(x\sqrt{\Omega_k}\big)\big/\sqrt{\Omega_k}, &\Omega_k>0,\\
 x, & \Omega_k=0,\\ \sin\big(x\sqrt{|\Omega_k|}\big)\big/\sqrt{|\Omega_k|}, &
 \Omega_k<0;
 \end{array}\right.$$
redshift $z$ and  the Hubble parameter $H(z)$ are connected with the scale factor
$a(t)$:
 \begin{equation}
 a(t)=\frac{a_0}{1+z},
\qquad
 H(z)=\frac{\dot a(t)}{a(t)};
 \label{Hz} \end{equation}
 $k$ is the sign of curvature, $\Omega_k=-k\big/(a_0^2H_0^2)$
 is its present time fraction, $a_0\equiv a(t_0)$ and $H_0\equiv H(t_0)$ are the current
 values of $a$ and $H$.

For any cosmological model we fix its model parameters $p_1,p_2,\dots$, calculate
dependence $a(t)$, the integral (\ref{DL})
 and this model predicts theoretical values $D_L^{th}$ for
luminosity distance  (\ref{DL}) (for given $z$), or $\mu^{th}$ for modulus. To compare
these theoretical values with the observational data $z_i$ and $\mu_i^{obs}$ from the
table  \cite{SNTable} we use the function
 \begin{equation}
 \chi^2_{SN}(p_1,p_2,\dots)=\min_{H_0}\sum_{i,j=1}^{N_{SN}}
 \Delta\mu_i\big(C_{SN}^{-1}\big)_{ij} \Delta\mu_j,\qquad
 \Delta\mu_i=\mu^{th}(z_i,H_0,p_1,\dots)-\mu_i^{obs}.
  \label{chiS} \end{equation}
Here $C_{SN}$ is the SN-by-SN covariance matrix \cite{SNTable}, representing systematic
errors.

In the sum (\ref{DL}) marginalization over the Hubble constant $H_0$ is assumed, because
we have to take into account  model dependence of the moduli $\mu_i^{obs}$. Unlike
observed apparent magnitudes $m_i^{obs}$  the values $\mu_i^{obs}$ in
Ref.~\cite{SNTable} are estimated as
 \begin{equation}
  \mu^{obs}=m^{obs}(z)-M+\alpha x_1-\beta c+\delta P.
  \label{muobs} \end{equation}
 This formula includes  the SN Ia absolute magnitude $M$  and corrections connected with
deviations from mean values of lightcurve shape ($x_1$), 
SN Ia color ($c$) and mass of a host galaxy (the factor $P$). The parameters $M$,
$\alpha$, $\beta$ and $\delta$ are considered in Ref.~\cite{SNTable} as nuisance
parameters, they are fitted simultaneously with the cosmological parameters in the flat
$\Lambda$CDM model.

Thus we have a model dependent additive term in Eq.~(\ref{muobs}) for the Union2.1
values $\mu^{obs}$ \cite{SNTable} with concealed dependence on the Hubble constant $H_0$
and other model parameters. In particular, one can find only that the best fit value for
the absolute magnitude $M=-19.321\pm0.03$ is obtained in Ref.~\cite{SNTable} for
$h=0.7$, where $h=H_0/100$ km\,s${}^{-1}$Mpc${}^{-1}$.

To suppress this dependence many authors \cite{NessPeri05,FarooqMR13,FarooqR13,ShiHL12}
suppose that values $\mu_i^{obs}$ from any SN Ia survey have a 
systematic error depending on $H_0$ and marginalize the sum (\ref{DL}) over the Hubble
constant $H_0$. They use the fact, that for the most popular models theoretical value of
the luminosity distance (\ref{DL}) depends on $H_0$ as $D_L^{th}\sim H_0^{-1}$, so the
distance modulus $\mu^{th}$ has the additive term $-5\log_{10}H_0$. In
Ref.~\cite{NessPeri05} this term is separated as $\mu_0$ in the form
 $$\mu(D_L)=5\log_{10}\frac{H_0D_L}c+\mu_0,\qquad \mu_0=42{.}384-5\log_{10}h;$$
 if we denote $d_i= \Delta\mu_i-\mu_0$, $\mathbf{d}=(d_1,\dots,d_{SN})$,
the minimum of the sum (\ref{chiS}) over $H_0$ (or over $\mu_0$) will take the form
  \begin{equation}
\chi^2_{SN}(p_1,\dots)=\mathbf{d} C_{SN}^{-1}\mathbf{d}^{T}- \frac{B^2}C,\qquad
B=\sum_{i,j=1}^{N_{SN}} d_i\big(C_{SN}^{-1}\big)_{ij},\;\quad C=\sum_{i,j=1}^{N_{SN}}
\big(C_{SN}^{-1}\big)_{ij}.
  \label{chiSm} \end{equation}

In this paper for all models we use the marginalized function (\ref{chiSm})
$\chi^2_{SN}$ to describe the supernovae Ia data \cite{SNTable}.

This approach with separation of the Hubble constant $H_0$ among other model parameters
can not be applied to $H(z)$ and BAO observational data, because observed values have
different dependence on $H_0$.

To describe the BAO data we 
calculate the distance \cite{Eisen05,WMAP,Plank13,Plank15}
 \begin{equation}
D_V(z)=\bigg[\frac{cz D_L^2(z)}{(1+z)^2H(z)}\bigg]^{1/3},
 \label{DV} \end{equation}
and  two measured  values
 \begin{equation}
 d_z(z)= \frac{r_s(z_d)}{D_V(z)},\qquad
  A(z) = \frac{H_0\sqrt{\Omega_m}}{cz}D_V(z),
  \label{dzAz} \end{equation}
which are usually considered as observational manifestations of baryon acoustic
oscillations \cite{Eisen05,WMAP}. Here
  $\Omega_m=\frac83\pi G\rho(t_0)\big/H_0^2$
is the present time fraction of matter with density $\rho$. The value $r_s(z_d)$ in
Eq.~(\ref{dzAz}) is sound horizon size at the end of the drag era $z_d$:
 \begin{equation}
r_s(z_d)= \int_{z_d}^\infty\frac{c_s(z)}{H(z)}dz,
  \label{rszd} \end{equation}
 To estimate this important parameter different authors
\cite{Blake12,Busca12,ChuangW12,Chuang13,Anderson13,Anderson14,Oka13,Font-Ribera13,Delubac14,Percival09,Kazin09,Beutler11,BlakeBAO11,Padmanabhan12,Seo12,Kazin14,Ross14}
used theoretical or statistical approaches and suggested different fitting formulas for
$r_s$.
 In table~\ref{rd}  the following recent estimations of $r_s(z_d)\equiv r_d$ and  $r_d h$ are shown:
\begin{table}[h]
 {\begin{tabular}{||c||c|c|c|c|c|c|c|c|c|c|c||}
\hline
 Refs &
\cite{Delubac14}& \cite{Plank13,Font-Ribera13} &\cite{Plank15}
&\cite{Kazin14}&\cite{Ross14} &\cite{Anderson14}  &\cite{BlakeBAO11}
 &\cite{Busca12}&\cite{Anderson13} &  \cite{Kazin09,Beutler11,Seo12}  &\cite{Percival09} \\ 
 \hline
  $ r_s(z_d) $ &   $147.4$   &   $147.49$  &  $147.6$ & $148.6$ &  $148.69$  &  $149.28$  &  $152.40$  &
    $152.76$  &  $153.19$  & $153.2$ & $153.5$ \\
  \hline
  $r_d\cdot h$ &   $98.79$   &  $99.26$ &   $100.09$  & $99.78$ & $99.62$ &  $104.49$  &  $105.15$  &  $106.93$  &
    $107.23$  & $103.6$ &  $104.69$  \\
 \hline
 \end{tabular}
 \caption{Recent estimations of $r_s(z_d)$ and $r_s(z_d)\cdot h$ (Mpc).}
\label{rd}}\end{table}

In the most of cited papers in table~\ref{rd} the values $r_s(z_d)$ were considered as
fiducial ones for calculating $D_V(z)$, $H(z)$ and other parameters. So these results
sufficiently depend on $r_s(z_d)$, in particular, estimations of $H(z)$ from the BAO
data
\cite{Gazta09,Blake12,Busca12,ChuangW12,Chuang13,Anderson13,Anderson14,Oka13,Font-Ribera13,Delubac14}
are defined with the factor $r_d^{fid}/r_d$, the corresponding factor $r_d/r_d^{fid}$
takes place for calculated values $D_V(z)$ or $D_A(z)=D_L(z)/(1+z)^2$.

In this paper we use two different approaches to calculate the sound horizon scale
$r_s(z_d)$. But previously one should mention the simplest method, applied in
Ref.~\cite{ShV14}, where the arithmetic average of the $r_d$ values in table~\ref{rd}
(with their multiplicity)
 \begin{equation}
 r_s(z_d)=150.69\pm 2.45\mbox{ Mpc}
 \label{rs} \end{equation}
was used as the basic value. This value is independent on $H_0$, hence the observational
parameter $d_z(z)$ in Eq.~(\ref{dzAz}) appears to be Hubble dependent (though the
formula (\ref{rszd}) predicts $ r_d\sim H_0^{-1}$ and Hubble free  $d_z$). One may
conclude, that $h$ dependence of $d_z$ is the drawback of this approach, so in this
paper we consider the fixed value $r_d$ (\ref{rs})  only in section~\ref{LCDM} to
emphasize advantages of other methods.

More appropriate procedures to calculate $r_s(z_d)$ include different fitting formulae
\cite{EisenHu98,Aubourg14,Anderson14,ShiHL12}. In this paper we use the numerically
calibrated approximation from Ref.~\cite{Aubourg14}
 \begin{equation}
 r_s(z_d)=\frac{55.154\exp\big[72.3(\Omega_\nu h^2 + 0.0006)^2\big]}
{(\Omega_m h^2)^{0.25351} (\Omega_b h^2)^{0.12807}}\mbox{ Mpc}
 \label{rsA} \end{equation}
 as the basic formula. The resulting $h$ dependence in Eq.~(\ref{rsA})
 (for a reasonable neutrino contribution with $\sum m_\nu \le 0.23$ eV \cite{Plank15})
is $r_d\sim h^{-0.7632}$, it is more close to the true variant $r_d\sim h^{-1}$. The
dependence on $\Omega_m$ in Eq.~(\ref{rsA}) is well fitted for  $\Lambda$CDM-like
models, however for the models with Chaplygin gas and with quadratic EoS, considered
below, $\Omega_m$ is not a basic model parameter. The value $\Omega_m$ in these models
should be estimated in a special way, so an additional uncertainty appears in this
approach.

Thus, an alternative simple fitting formula
 \begin{equation}
 r_s(z_d)=\frac{(r_d\cdot h)_{fid}}h,\qquad
(r_d\cdot h)_{fid}=104.57\mbox{ Mpc}.
 \label{rsh}\end{equation}
 with true $h$ dependence may be suggested. Here the value $(r_d\cdot h)_{fid}=104.57\pm1.44$ Mpc was
chosen as the best fit for the $\Lambda$CDM model. This procedure is described in the
next section and illustrated in figure~\ref{F0}.

The parameter $(r_d\cdot h)_{fid}$ for the expression (\ref{rsh}) plays the same role as
the baryonic fraction $\Omega_b$ for the formula~(\ref{rsA}), in both cases we do not
consider $\Omega_b$ and $r_d h$ as free model parameters for all models, but fix their
optimal (fiducial) values after description of the simplest $\Lambda$CDM model. The best
$\Lambda$CDM fit for $\Omega_b$ in Eq.~(\ref{rsA}) (see  figure~\ref{F0}) is
 \begin{equation}
 \Omega_b=0.044\pm0.004.
 \label{Omb} \end{equation}

\medskip

To take into account all available BAO data
\cite{Gazta09,Blake12,Busca12,ChuangW12,Chuang13,Anderson13,Anderson14,Oka13,Font-Ribera13,Delubac14,Percival09,Kazin09,Beutler11,BlakeBAO11,Padmanabhan12,Seo12,Kazin14,Ross14}
for parameters (\ref{dzAz}), we consider in this paper $N_{BAO}=17$ data points for
$d_z(z)$ (10 additional points in comparison with the table in our paper \cite{ShV14})
and 7 data points for $A(z)$ presented in the table~\ref{TBAO}.

\begin{table}[th]
\centering
 {\begin{tabular}{||l|l|l|l|l|c|l||}
\hline
 $z$  & $d_z(z)$ &$\sigma_d$    & ${ A}(z)$ & $\sigma_A$  & Refs & Survey\\ \hline
 0.106& 0.336  & 0.015 & 0.526& 0.028& \cite{WMAP,Beutler11}  & 6dFGS \\ \hline
 0.15 & 0.2232 & 0.0084& -    & -    & \cite{Ross14}& SDSS DR7  \\ \hline
 0.20 & 0.1905 & 0.0061& 0.488& 0.016& \cite{Percival09,BlakeBAO11}  & SDSS DR7 \\ \hline
 0.275& 0.1390 & 0.0037& -    & -    & \cite{Percival09}& SDSS DR7 \\ \hline
 0.278& 0.1394 & 0.0049& -    & -    & \cite{Kazin09}  &SDSS DR7 \\ \hline
 0.314& 0.1239 & 0.0033& -    & -    & \cite{BlakeBAO11}& SDSS LRG \\ \hline
 0.32 & 0.1181 & 0.0026& -    & -    & \cite{Anderson14} &BOSS DR11 \\ \hline
 0.35 & 0.1097 & 0.0036& 0.484& 0.016& \cite{Percival09,BlakeBAO11} &SDSS DR7 \\ \hline
 0.35 & 0.1126 & 0.0022& -    & -    & \cite{Padmanabhan12}   &SDSS DR7 \\ \hline
 0.35 & 0.1161 & 0.0146& -    & -    & \cite{ChuangW12}   &SDSS DR7 \\ \hline
 0.44 & 0.0916 & 0.0071& 0.474& 0.034& \cite{BlakeBAO11}& WiggleZ \\ \hline
 0.57 & 0.0739 & 0.0043& 0.436& 0.017& \cite{Chuang13}& SDSS DR9 \\ \hline
 0.57 & 0.0726 & 0.0014& -    & -    & \cite{Anderson14}& SDSS DR11 \\ \hline
 0.60 & 0.0726 & 0.0034& 0.442& 0.020& \cite{BlakeBAO11} & WiggleZ \\ \hline
 0.73 & 0.0592 & 0.0032& 0.424& 0.021& \cite{BlakeBAO11} &WiggleZ \\ \hline
 2.34 & 0.0320 & 0.0021& -& - & \cite{Delubac14} & BOSS DR11 \\ \hline
 2.36 & 0.0329 & 0.0017& -& - & \cite{Font-Ribera13} & BOSS DR11 \\  \hline
 \end{tabular}
 \caption{Values of $d_z(z)=r_s(z_d)/D_V(z)$ and $A(z)$ (\ref{dzAz})
with errors and references} \label{TBAO}}\end{table}

Measurements of $d_z(z)$ and $A(z)$ from Refs.~\cite{Percival09,BlakeBAO11} in
table~\ref{TBAO}
 are not independent. So the $\chi^2$ function for
the values (\ref{dzAz}) is
 \begin{equation}
 \chi^2_{BAO}(p_1,p_2,\dots)=(\Delta d)^TC_d^{-1}\Delta d+
(\Delta { A})^TC_A^{-1}\Delta  A,\qquad \Delta d=d_z(z_i)-d_z^{th}.
  \label{chiB} \end{equation}
 The elements of covariance matrices
$C_d^{-1}=||c^d_{ij}||$ and $C_A^{-1}=||c^A_{ij}||$ in Eq.~(\ref{chiB}) are
\cite{WMAP,Percival09,BlakeBAO11}:
$$\begin{array}{llll}
c^d_{33}=30124,& c^d_{38}=-17227,&  c^d_{88}=86977,& \\
c^d_{1\!11\!1}=24532.1,& c^d_{1\!11\!4}=-25137.7,&  c^d_{1\!11\!5}=12099.1,& c^d_{1\!41\!4}=134598.4,\\
 c^d_{1\!41\!5}=-64783.9,& c^d_{1\!51\!5}=128837.6; & c^A_{1\!11\!1}=1040.3,& c^A_{1\!11\!4}=-807.5,\\
 c^A_{1\!11\!5}=336.8, & c^A_{1\!41\!4}=3720.3,& c^A_{1\!41\!5}=-1551.9,& c^A_{1\!51\!5}=2914.9.
 \end{array}$$
 Here $c_{ij}=c_{ji}$, the remaining matrix elements are
 $c_{ii}=1/\sigma_i^2$, $c_{ij}=0$, $i\ne j$.

In the values $\sigma_d$ in table~\ref{TBAO} we took into account correlation between
estimations of $d_z(z)$ and $H(z)$ (table~\ref{TH}) for $z=0.35$, 0.57,  2.34,  2.36  in
Refs.~\cite{ChuangW12,Chuang13,Anderson13,Font-Ribera13,Delubac14}.

\medskip

Measurements of the Hubble parameter $H(z)$ for different redshifts $z$ with 38 data
points
\cite{Simon05,Stern10,Moresco12,Zhang12,Moresco15,Gazta09,Blake12,Busca12,ChuangW12,Chuang13,Anderson13,Anderson14,Oka13,Font-Ribera13,Delubac14}
are presented in table~\ref{TH}.
  These values $H(z)$ were calculated with two methods:
1) differential age approach  in
Refs.~\cite{Simon05,Stern10,Moresco12,Zhang12,Moresco15} with evaluation of the age
difference $dt$ for galaxies with close redshifts $dz$ and the formula
 $$  H(z)=\frac1{a(t)}\frac{da}{dt}=-\frac1{1+z}\frac{dz}{dt},
 $$
2) measurement
\cite{Gazta09,Blake12,Busca12,ChuangW12,Chuang13,Anderson13,Anderson14,Oka13,Font-Ribera13,Delubac14}
 of the BAO peak in the correlation function in
line-of-sight directions at a redshift separation $\Delta z = r_s(z_d)\,H(z)/c$.

\begin{table}[h]
\centering
 {\begin{tabular}{||l|l|l|c||l|l|l|c||}   \hline
 $z$  & $H(z)$ &$\sigma_H$  & Refs  &   $z$ & $H(z)$ & $\sigma_H$  & Refs\\ \hline
 0.070 & 69  & 19.6& \cite{Zhang12}  & 0.570 & 96.8& 3.4 & \cite{Anderson14}\\ \hline
 0.090 & 69  & 12  & \cite{Simon05}  & 0.593 & 104 & 13  & \cite{Moresco12} \\ \hline
 0.120 & 68.6& 26.2& \cite{Zhang12}  & 0.600 & 87.9& 6.1 & \cite{Blake12}   \\ \hline
 0.170 & 83  & 8   & \cite{Simon05}  & 0.680 & 92  & 8   & \cite{Moresco12} \\ \hline
 0.179 & 75  & 4   & \cite{Moresco12}& 0.730 & 97.3& 7.0 & \cite{Blake12}  \\ \hline
 0.199 & 75  & 5   & \cite{Moresco12}& 0.781 & 105 & 12  & \cite{Moresco12}\\ \hline
 0.200 & 72.9& 29.6& \cite{Zhang12}  & 0.875 & 125 & 17  & \cite{Moresco12}\\ \hline
 0.240 &79.69& 2.99& \cite{Gazta09}  & 0.880 & 90  & 40  & \cite{Stern10}  \\ \hline
 0.270 & 77  & 14  & \cite{Simon05}  & 0.900 & 117 & 23  & \cite{Simon05}  \\ \hline
 0.280 & 88.8& 36.6& \cite{Zhang12}  & 1.037 & 154 & 20  & \cite{Moresco12}\\ \hline
 0.300 & 81.7& 6.22& \cite{Oka13}    & 1.300 & 168 & 17  & \cite{Simon05}  \\ \hline
 0.340 & 83.8& 3.66& \cite{Gazta09}  & 1.363 & 160 & 33.6& \cite{Moresco15}\\ \hline
 0.350 & 82.7& 9.1 & \cite{ChuangW12}& 1.430 & 177 & 18  & \cite{Simon05}  \\ \hline
 0.352 & 83  & 14  & \cite{Moresco12}& 1.530 & 140 & 14  & \cite{Simon05}  \\ \hline
 0.400 & 95  & 17  & \cite{Simon05}  & 1.750 & 202 & 40  & \cite{Simon05}  \\ \hline
 0.430 &86.45& 3.97& \cite{Gazta09}  & 1.965 &186.5& 50.4& \cite{Moresco15}\\ \hline
 0.440 & 82.6& 7.8 & \cite{Blake12}  & 2.300 & 224 & 8.6 & \cite{Busca12}  \\ \hline
 0.480 & 97  & 62  & \cite{Stern10}  & 2.340 & 222 & 8.5 & \cite{Delubac14}\\ \hline
 0.570 & 87.6& 7.8 & \cite{Chuang13} & 2.360 & 226 & 9.3 & \cite{Font-Ribera13}\\  \hline
 \end{tabular}
\caption{Values of the Hubble parameter $H(z)$ with errors $\sigma_H$ from Refs.~
\cite{Simon05,Stern10,Moresco12,Zhang12,Moresco15,Gazta09,Blake12,Busca12,ChuangW12,Chuang13,Anderson13,Anderson14,Oka13,Font-Ribera13,Delubac14}}
 \label{TH}}\end{table}

For the latter method estimations
\cite{Gazta09,Blake12,Busca12,ChuangW12,Chuang13,Anderson13,Anderson14,Oka13,Font-Ribera13,Delubac14}
of $H(z)$ essentially depend on a fiducial value $r_d^{fid}$ and have the factor
$r_d^{fid}/r_d$, as was mentioned above. In particular, the result in
Ref.~\cite{Delubac14} is
 $$H(z = 2.34) = \big(222\pm 7 \big)\,\frac{\mbox{km}}{\mbox{s}\cdot\mbox{Mpc}}\cdot\frac{147.4\mbox{ Mpc}}{r_s(z_d)}.$$
 In table~\ref{TH} this factor is taken into account only for the errors $\sigma_H$
from the papers \cite{Gazta09,Busca12,ChuangW12,Chuang13,Font-Ribera13,Delubac14}, where
fiducial values $r_d^{fid}$ essentially differ from the average (\ref{rs}). The
estimations for  $H(z)$ in table~\ref{TH} are the same as in the correspondent sources.

To compare the $H(z)$ data in table~\ref{TH} with $N_H=38$ data points with model
predictions we use the $\chi^2$ function
\begin{equation}
 \chi^2_H(p_1,p_2,\dots)=\sum_{i=1}^{N_H}
 \frac{\big[H_i-H^{th}(z_i,p_1,p_2,\dots)\big]^2}{\sigma_{H,i}^2},
  \label{chiH} \end{equation}
similar to the function (\ref{chiS}) for the SN Ia observational data from
Ref.~\cite{SNTable}.

We mentioned above that the observed values in tables~\ref{TBAO},~\ref{TH} have
different dependence on $h=H_0/100$ km\,s${}^{-1}$Mpc${}^{-1}$. In particular,
if we use the fitting formula (\ref{rsA}) for $r_s(z_d)$, the parameter $d_z$
(\ref{dzAz}) has rather weak  $h$ dependence ($d_z\sim  h^{0.2368}$,  because $D_V\sim
h^{-1}$); for the formula (\ref{rsh}) $d_z$ is Hubble free in accordance with
Eq.~(\ref{dzAz}). The value $A(z)$ is also Hubble free, but it depends on $\Omega_m$.
The values $H(z)$ in table~\ref{TH} are naturally proportional to $H_0$. So estimations
in tables~\ref{TBAO},~\ref{TH} may be model dependent, but we can assume that different
authors use different methods and produce possible systematic errors for $d_z(z)$ and
$H(z)$ in different directions. One can suppose mean systematic errors to be close to
zero.

On the other hand, if any form of marginalization over  $H_0$ for BAO and  $H(z)$ data
is made \cite{FarooqMR13,FarooqR13,ShiHL12}, the obtained results will have an
additional error, because a model can successfully describe all SN Ia, BAO and  $H(z)$
data, but with 3 essentially different intrinsic values of $H_0$. Under these arguments
we make the marginalization procedure (\ref{chiSm}) over  $H_0$ only for SN Ia data
\cite{SNTable}, but not for BAO and $H(z)$ data from  tables~\ref{TBAO},~\ref{TH}.

\section{$\Lambda$CDM model
}\label{LCDM}


In the $\Lambda$CDM and other models  
in this paper the Einstein equations
 \begin{equation}
 G^\mu_\nu=8\pi G T^\mu_\nu-\Lambda\delta^\mu_\nu
 \label{Eeq}\end{equation}
  describe dynamics of the universe. Here
$G^\mu_\nu=R^\mu_\nu-\frac12R\delta^\mu_\nu$, $ T^{\mu}_{\nu} =
\mbox{diag}\,(-\rho,p,p,p)$.

 In the $\Lambda$CDM model
baryonic and dark matter may be described as one component of dust-like matter with
density $\rho=\rho_c=\rho_b+\rho_{dm}$, so we suppose $p=0$ in $ T^{\mu}_{\nu}$. In
models with Chaplygin gas (Sect.~\ref{Chap}) and with quadratic equation of state
(Sect.~\ref{Quad}) we suppose that an additional component of matter describes both dark
matter and dark energy and gives some contribution $\rho_g$ in the total density:
 \begin{equation}
 \rho=\rho_c+\rho_g+\rho_r.
 \label{rho}
  \end{equation}
The fraction of relativistic matter (radiation and neutrinos) is close to zero for
observable values $z\le2.36$, so below we suppose $\rho_r=0$.

 For the Robertson-Walker metric with the curvature sign $k$
 \begin{equation}
 ds^2 = -dt^2+a^2(t)\Big[(1-k r^2)^{-1}dr^2+r^2
 d\Omega\Big]
 \label{metrRW}
 \end{equation}
 the Einstein equations (\ref{Eeq}) are reduced to the system
\begin{eqnarray}
3\frac{\dot{a}^2+k}{a^2}=8\pi G\rho+\Lambda,\label{Esysa}\\
 \dot{\rho}=-3\frac{\dot{a}}{a}(\rho+p).\label{Esys2}
\end{eqnarray}
 Eq.~(\ref{Esys2}) results from the continuity
condition $T^{\mu}_{\nu; \mu}=0$, the dot denotes the time derivative, here and below
the speed of light $c=1$.


For the $\Lambda$CDM with dust-like matter ($p=0$) we use the solution
 of Eq.~(\ref{Esys2})  $\rho/\rho_0=(a/a_0)^{-3}$ and
rewrite Eq.~(\ref{Esysa}) in the form
 \begin{equation}
 \frac{\dot{a}^2}{a^2H_0^2}=\frac{H^2}{H_0^2}=\Omega_m \Big(\frac{a}{a_0}\Big)^{-3}+
 \Omega_\Lambda+\Omega_k\Big(\frac{a}{a_0}\Big)^{-2}.
  \label{EqLCDM} \end{equation}
 Here the present time fractions of matter, dark energy ($\Lambda$ term) and curvature
 \begin{equation}
\Omega_m
 =\frac{8\pi G\rho(t_0)}{3H_0^2},\qquad
\Omega_\Lambda=\frac{\Lambda}{3H_0^2},\qquad
 \Omega_k=-\frac{k}{a_0^2H_0^2}
 \label{Omega1} \end{equation}
 are  connected by the equality
 \begin{equation}
\Omega_m+\Omega_\Lambda+ \Omega_k=1,
 \label{sumOm} \end{equation}
resulting from Eq.~(\ref{EqLCDM}) if we fix $t=t_0$.

If we introduce  dimensionless time $\tau$ and logarithm of the scale factor
\cite{GrSh13}
 \begin{equation}
\tau=H_0t,\qquad
 {\cal A}=\log\frac a{a_0}.
 \label{tau} \end{equation}
 equation (\ref{EqLCDM}) will take the form \ 
 $\frac{d{\cal A}}{d\tau}=\sqrt{\Omega_m e^{-3{\cal A}}+
 \Omega_\Lambda+\Omega_ke^{-2{\cal A}}}$,
 more convenient for numerical solving with the initial
condition at the present time ${\cal A}\big|_{\tau=1}=0$ equivalent to $a(t_0)=a_0$.
Here and below  the present time $t=t_0$ corresponds to $\tau=1$.

If we fix all model parameters, we can solve numerically the Cauchy problem for
Eq.~(\ref{EqLCDM}) and calculate the values $a(t)/a_0$, $H(z)$, $D_L(z)$ (\ref{DL}),
$d_z(z)$ and $A(z)$ (\ref{dzAz}). To compare them with the observational data from
Ref.~\cite{SNTable} and tables~\ref{TBAO}, \ref{TH} we use the $\chi^2$ functions
(\ref{chiS}), (\ref{chiB}) and (\ref{chiH})
and (under assumption about their Gaussian nature) the correspondent summarized function
  \begin{equation}
\chi^2_\Sigma=\chi^2_{SN}+\chi^2_H+\chi^2_{BAO}.
 \label{chisum} \end{equation}

When we apply the $\Lambda$CDM model for describing the  observational data from
Sect.~\ref{Observ} (for $z\le2.36$), we use three free model parameters  $H_0$,
$\Omega_m$ and $\Omega_\Lambda$ (or $\Omega_k$ instead of  $\Omega_\Lambda$) and the
additional parameter
 \begin{equation}
 \Omega_b=\frac{\rho_b(t_0)}{\rho_{cr}}
 =\frac{8\pi G\rho_b(t_0)}{3H_0^2},
 \label{Omegab} \end{equation}
 if we use the fitting formula (\ref{rsA}) for $r_s(z_d)$. For the formula (\ref{rsh})
the baryonic fraction $\Omega_b$ is not a model parameter, we mentioned above that the
value $r_d\cdot h$ plays the role of an additional parameter in the case (\ref{rsh}). In
both approaches we test the $\Lambda$CDM model and estimate the best $\Lambda$CDM fit
correspondingly for $\Omega_b$ and $r_d\cdot h$.

\begin{figure}[th]
  \centerline{\includegraphics[scale=0.72,trim=7mm 2mm 2mm 2mm]{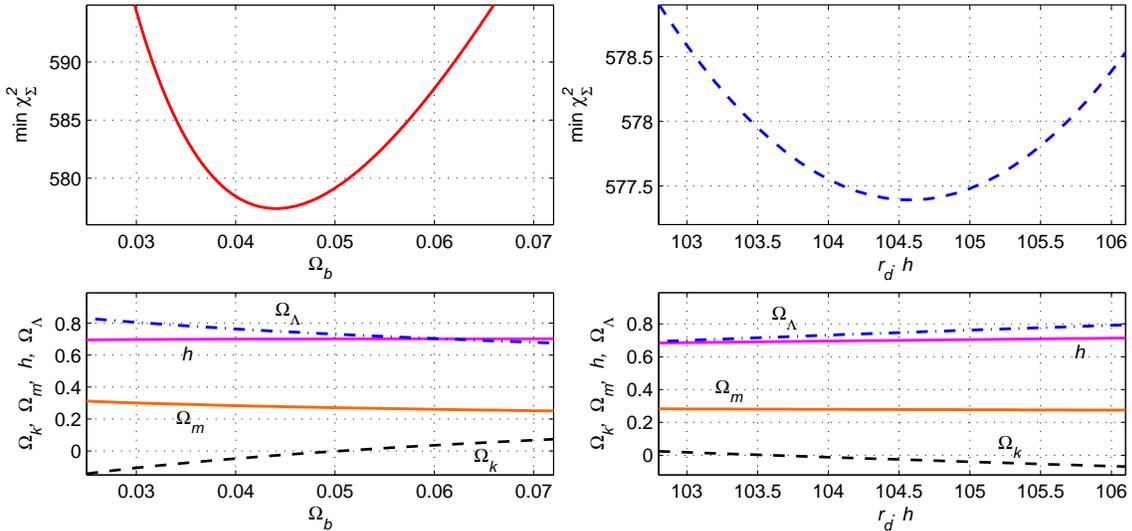}}
 \caption{\small  The $\Lambda$CDM model: dependence of the best fit
$\min\limits_{H_0,\Omega_m,\Omega_\Lambda}\chi^2_\Sigma$ on
 $\Omega_b$ for the fitting formula (\ref{rsA}) (the left panels) and on
 $r_d\cdot h$ for  Eq.~(\ref{rsh}) (the right panels). In the bottom panels we show
the correspondent dependencies for parameters $H_0$, $\Omega_m$, $\Omega_\Lambda$,
$\Omega_k$ of the $\chi^2_\Sigma$ minimum point. }
  \label{F0}
\end{figure}

The results are presented in figure~\ref{F0}, where dependence of the best (minimal)
value of the function (\ref{chisum})
$\min\chi^2_\Sigma=\min\limits_{H_0,\Omega_m,\Omega_\Lambda}\chi^2_\Sigma$ on $\Omega_b$
is shown in the top-left panel for the case (\ref{rsA}). Here and for all models we
assume $\sum m_\nu = 0.06$  eV  \cite{Aubourg14,Plank13}. One should note that the
formula (\ref{rsA}) is insensitive to a neutrino contribution in the range $\sum m_\nu
\le 0.23$ eV \cite{Plank15}.

For the variant with Eq.~(\ref{rsh}) the similar dependence of $\min\chi^2_\Sigma$ on
$r_d\cdot h$ is in the top-right panel. Here and below we draw these graphs as solid red
lines for the fitting formula (\ref{rsA}) and as dashed blue lines for the variant
(\ref{rsh}).

The bottom panels of figure~\ref{F0} illustrate how coordinates $h$, $\Omega_m$,
$\Omega_\Lambda$ and $\Omega_k=1-\Omega_m-\Omega_\Lambda$ of the minimum point for
$\min\limits_{H_0,\Omega_m,\Omega_\Lambda}\chi^2_\Sigma$ depend on the correspondent
parameters $\Omega_b$ and $r_d\cdot h$.

One can see that for the formula (\ref{rsA}) dependence of $\min\chi^2_\Sigma$ on
$\Omega_b$ is rather sharp: we have the distinct minimum at the value  (\ref{Omb})
$\Omega_b=0.044$. Below we use this value as the fiducial one for all models. The
correspondent dependence in the top-right panel is more smooth, however it results in
the optimal (fiducial) value  $r_d\cdot h=104.57\pm1.44$ Mpc in the formula (\ref{rsh}).

\begin{figure}[th]
  \centerline{\includegraphics[scale=0.72,trim=7mm 2mm 2mm 2mm]{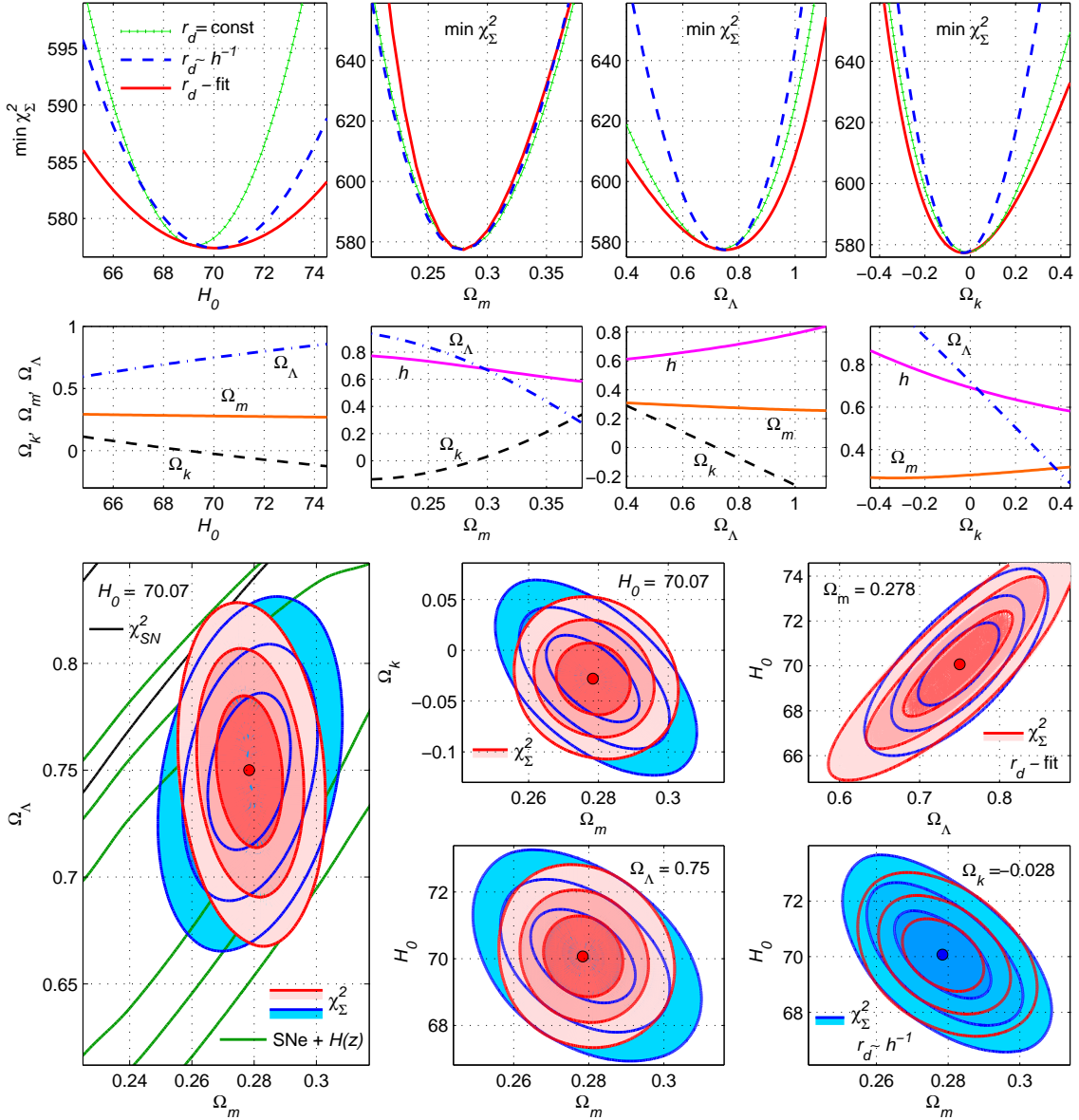}}
  \caption{\small For the $\Lambda$CDM in the top panels we present how $\min\chi^2_\Sigma$
depends on $H_0$, $\Omega_m$, $\Omega_\Lambda$ and  $\Omega_k$ with different fitting
formulae for $r_d$: for Eq.~(\ref{rsA}) with $\Omega_b=0.044$ (red solid lines), for
Eq.~(\ref{rsh}) (blue dashed lines), for Eq.~(\ref{rs}) (green lines with dots). The
correspondent dependencies for parameters of $\chi^2_\Sigma$ minimum point for
Eq.~(\ref{rsA}) are shown in the 2-nd row. In other panels $1\sigma$, $2\sigma$ and
$3\sigma$ level lines are drawn for $\chi^2_\Sigma$ as red filled contours for
Eq.~(\ref{rsA}) and  blue filled contours for Eq.~(\ref{rsh}) and also as black lines
for $\chi^2_{SN}$ and green lines for $\chi^2_{SN}+\chi^2_H$; optimal values of the
third parameter are fixed and shown. }
  \label{F1}
\end{figure}

If we fix these parameters as described above, we can test the  $\Lambda$CDM model for
different values of the remaining 3 parameters: $H_0$, $\Omega_m$ and $\Omega_\Lambda$.
The results of calculations are presented in tables~\ref{WMPl}, \ref{Estim},
\ref{Estimrh} and in  figure~\ref{F1}. In the top-left panel of the figure we see how
minimum of the function (\ref{chisum})
$\min\chi^2_\Sigma=\min\limits_{\Omega_m,\Omega_\Lambda}\chi^2_\Sigma(H_0)$ depend on
the Hubble constant $H_0$: red solid lines in the top panels describe the model with the
formula (\ref{rsA}), blue dashed lines correspond to the variant (\ref{rsh}). For the
sake of comparison we present here graphs for the fixed value  $r_s(z_d)$ (\ref{rs}) as
green lines with dots.  These minima are calculated for each fixed value $H_0$.

One may see that the function $\min\chi^2_\Sigma(H_0)$ for the fitting formula
(\ref{rsA}) (the red line) has the maximal spread and achieves its minimum
$\min\chi^2_\Sigma\simeq577.39$ at
 \begin{equation}
 H_0=70.07\pm1.82\;\;
\mbox{ km\,(s\,Mpc)}^{-1}
 \label{H0LCDM} \end{equation}
 for the following values of other parameters: $\Omega_m\simeq0.278$, $\Omega_\Lambda\simeq0.75$
(presented in table~\ref{WMPl}). It is interesting to see in table~\ref{WMPl} and in
figure~\ref{F1} that for the fitting formula (\ref{rsh}) (blue dashed lines) the minimum
$\min\chi^2_\Sigma$ and the optimal values of all parameters are the same, there is some
difference only in $1\sigma$ errors. To estimate $1\sigma$ errors in the equality
(\ref{H0LCDM}) we use the one-dimensional likelihood function ${\cal
L}_\Sigma(H_0)\sim\exp(-\chi^2_\Sigma/2)$ corresponding to
$\min\limits_{\Omega_m,\Omega_\Lambda}\chi^2_\Sigma(H_0)$.

 Optimal values and errors for other model parameters
$\Omega_m$, $\Omega_\Lambda$, $\Omega_k$ are calculated similarly, they are tabulated
below in table~\ref{WMPl} in comparison with other estimates and also in the next
section in tables~\ref{Estim} and \ref{Estimrh} correspondingly for the expressions
(\ref{rsA}) and (\ref{rsh}).

Estimations of $\Omega_m$ and $\Omega_\Lambda$ in tables~\ref{WMPl}, \ref{Estim},
\ref{Estimrh} are connected with graphs in the next two panels in the top line of
figure~\ref{F1}, which present how minima of $\chi^2_\Sigma$ depend on $\Omega_m$ and on
$\Omega_\Lambda$. In particular, the solid red line in the second top panel describes
$\min\chi^2_\Sigma(\Omega_m)=\min\limits_{H_0,\Omega_\Lambda}\chi^2_\Sigma$; the
corresponding likelihood function ${\cal L}_\Sigma(\Omega_m)\sim\exp(-\chi^2_\Sigma/2)$
determines the $1\sigma$ error $\Delta\Omega_m\simeq0.008$ in
tables~\ref{WMPl},~\ref{Estim}. This panel demonstrates that dependencies of
$\min\chi^2_\Sigma$ on $\Omega_m$ for different variants of $r_s(z_d)$ are rather close
and have rather sharp form. It is connected with the contribution in $\chi^2_{BAO}$ from
the value $A(z)$ (\ref{dzAz}), because  $A(z)$ is proportional to $\sqrt{\Omega_m}$ and
$\chi^2_{BAO}$ is very sensitive to $\Omega_m$ values. For the red line in this panel we
have the additional dependence on $\Omega_m$ in the formula (\ref{rsA}).  In the third
panel the functions $\min\chi^2_\Sigma(\Omega_\Lambda)$ with distinct minima have some
difference in their the $1\sigma$ errors.

Estimations of $\Omega_k$ in tables~\ref{WMPl}\,--\,\ref{Estimrh} are calculated via the
function $\min\chi^2_\Sigma(\Omega_k)=\min\limits_{H_0,\Omega_m}\chi^2_\Sigma$. These
graphs are shown in the top-right panel of figure~\ref{F1}.

The panels in the second row in figure~\ref{F1} demonstrate how parameters of a minimum
point of the function  $\chi^2_\Sigma$ with the fitting formula (\ref{rsA}) depend on
$H_0$, $\Omega_m$, $\Omega_\Lambda$,  $\Omega_k$. In the left panel these coordinates
are $\Omega_m$ and $\Omega_\Lambda$, but also the value
$\Omega_k=1-\Omega_m-\Omega_\Lambda$ is drawn as the black dashed line. The optimal
value $\Omega_m$ remains practically constant in contrast with $\Omega_\Lambda$ and
$\Omega_k$. The graphs of $h(\Omega_m),\dots,h(\Omega_k)$ in other panels show
$h=H_0/100$, where $H_0$ is the optimal value corresponding to the  minimum point of
$\chi^2_\Sigma$.

In other panels of figure~\ref{F1} we present the results of calculations as level lines
at $1\sigma$ (68.27\%), $2\sigma$ (95.45\%) and $3\sigma$ (99.73\%) confidence levels
for the functions $\chi^2(p_1,p_2)$ in planes of two parameters, if the third parameter
is fixed. For example, the  functions $\chi^2_\Sigma(\Omega_m,\Omega_\Lambda)$ for the
fixed optimal value (\ref{H0LCDM}) of $H_0$ are shown in the bottom-left panel of
figure~\ref{F1} as red filled contours for the formula (\ref{rsA}) and  blue filled
contours for Eq.~(\ref{rsh}). The corresponding level lines for $\chi^2_{SN}$ and for
$\chi^2_{SN}+\chi^2_H$ are shown as black and green lines.

In other panels only $\chi^2_\Sigma$  filled contours for the cases (\ref{rsA}) and
(\ref{rsh}) are compared for different pairs of parameters. Points of minima for the
functions  $\chi^2_\Sigma$ are marked as red (or blue) circles, we mentioned above, that
they coincide for the variants (\ref{rsA}) and (\ref{rsh}). In two  bottom-left panels
the difference is in the fixed parameter ($\Omega_\Lambda$ or $\Omega_k$) and in a
choice of the foreground between the cases (\ref{rsA}) and (\ref{rsh}).


Our estimations of the $\Lambda$CDM parameters for two variants  (\ref{rsA}) and
(\ref{rsh}) of the fitting formula for $r_d$ are to be compared with the the following
best fits for these model parameters from surveys of the Wilkinson Microwave Anisotropy
Probe (WMAP) \cite{WMAP} and Planck Collaboration \cite{Plank13,Plank15} in
table~\ref{WMPl}.

 \begin{table}[h]
 {\centering
 \begin{tabular}{||l||c|c||c|c|c||} \hline
 & \multicolumn{2}{c||}{This paper } & \multicolumn{3}{c||}{WMAP, Planck surveys}\\
 \hline
 & $r_d$ (\ref{rsA}) & $r_d$ (\ref{rsh})  & WMAP 9y \cite{WMAP}&Planck\,13 \cite{Plank13} &Planck\,15 \cite{Plank15}  \\ \hline
 $H_0$ & $70.07\pm1.82$ & $70.07\pm1.27$ \rule{0pt}{1.2em}   & $69.7\pm2.4$ & $67.3\pm1.2$ & $67.8\pm0.9$ \\ \hline
 $\Omega_m$ & $0.278\pm 0.008$ & $0.278\pm 0.009$& $0.279\pm 0.025$ & $0.314\pm 0.02$ & $0.308\pm 0.012$ \\  \hline
 $\Omega_\Lambda$& $0.750_{-0.055}^{+0.051}$\rule{0pt}{1.2em}  & $0.750\pm0.034$& $0.721\pm 0.025$ & $0.686\pm 0.025$ & $0.692\pm 0.012$  \\  \hline
  $\Omega_k$ &$-0.028_{-0.048}^{+0.050}$ &  $-0.028_{-0.034}^{+0.035}$ & $-0.0027^{+0.0039}_{-0.0038}$& $-0.0005^{+0.0065}_{-0.0066}$ &
  $-0.005^{+0.016}_{-0.017}$\rule{0pt}{1.2em}  \\
 \hline
  $\Omega_b$ &$0.044\pm0.004$ &  - & $0.0463^{+0.0024}_{-0.0024}$& $0.0487^{+0.0018}_{-0.0017}$ &
  $0.0484^{+0.0014}_{-0.0013}$\rule{0pt}{1.2em}  \\
 \hline
 \end{tabular}
\caption{Estimations of the $\Lambda$CDM parameters.}
 \label{WMPl}}\end{table}

One can also add the estimates for the fixed $r_d$ (\ref{rs}): $H_0=69.27\pm0.93$
km\,s${}^{-1}$Mpc${}^{-1}$, $\Omega_m=0.280\pm 0.009$,
$\Omega_\Lambda=0.732_{-0.055}^{+0.044}$,  $\Omega_k=-0.012\pm0.045$ (corresponding to
green dots in the top panels in figure~\ref{F1}). In the case (\ref{rs})  the $1\sigma$
error for $H_0$ is smallest, because
 $d_z^{th}$ depends on $h$.

We see that our estimations of the model parameters in table~\ref{WMPl} for the cases
(\ref{rsA}) and (\ref{rsh}) are in good agreement with the WMAP estimates, but they are
in $1\sigma$ or $2\sigma$ tension with the values of Planck Collaboration. On the other
hand, all $H_0$ values in table~\ref{WMPl} have essential tension with the Hubble Space
Telescope group \cite{Riess11} estimation: $H_0=73.8\pm2.4$ km\,s${}^{-1}$Mpc${}^{-1}$.

The latter value  was used as a prior in Refs.~\cite{FarooqMR13,FarooqR13} for
describing Type Ia SNe, BAO and $H(z)$ data with the help of the $\Lambda$CDM, XCDM and
$\phi$CDM models. One may conclude, that for the $\Lambda$CDM this choice of $H_0$ was
unsuccessful in comparison with another value $H_0=68$ km\,s${}^{-1}$Mpc${}^{-1}$,
chosen in Refs.~\cite{FarooqMR13,FarooqR13}.

\section{Modified Chaplygin gas}
\label{Chap}

In the model with modified Chaplygin gas (MCG) this gas has the following equation of
state \cite{Benaoum02,Chimento04,DebnathBCh04,LiuLiC05,LuXuWL11,PaulTh13,PaulTh14}
 \begin{equation}
 p_g=w_0\rho_g-B\,\rho_g^{-\alpha}
 \label{EoSM}
  \end{equation}
for its  density $\rho_g$ as a part in the total density (\ref{rho}). MCG can unify dark
matter and dark energy. If $w_0=0$ the MCG model with Eq.~(\ref{EoSM}) is reduced to the
model with generalized Chaplygin gas  (GCG) with EoS \cite{KamenMP01,Bento02}
 \begin{equation}
p_g=-B\,\rho_g^{-\alpha}. \label{EoSGCG}
  \end{equation}
   In our paper \cite{ShV14}  the GCG
model was applied to describing the observational data for Type Ia supernovae
\cite{SNTable}, $H(z)$ with 34 data points and BAO with 7 data points for $d_z(z)$. In
this paper we consider the enlarged number of data points from tables~\ref{TBAO},
\ref{TH}, the more general  MCG model (\ref{EoSM}) (in comparison with the GCG case
$w_0=0$) and also we calculate the function  $\chi^2_{SN}$ (\ref{chiS}) with the
covariance matrix  $C_{SN}$.

\begin{figure}[th]
  \centerline{\includegraphics[scale=0.72,trim=5mm 0mm 5mm -1mm]{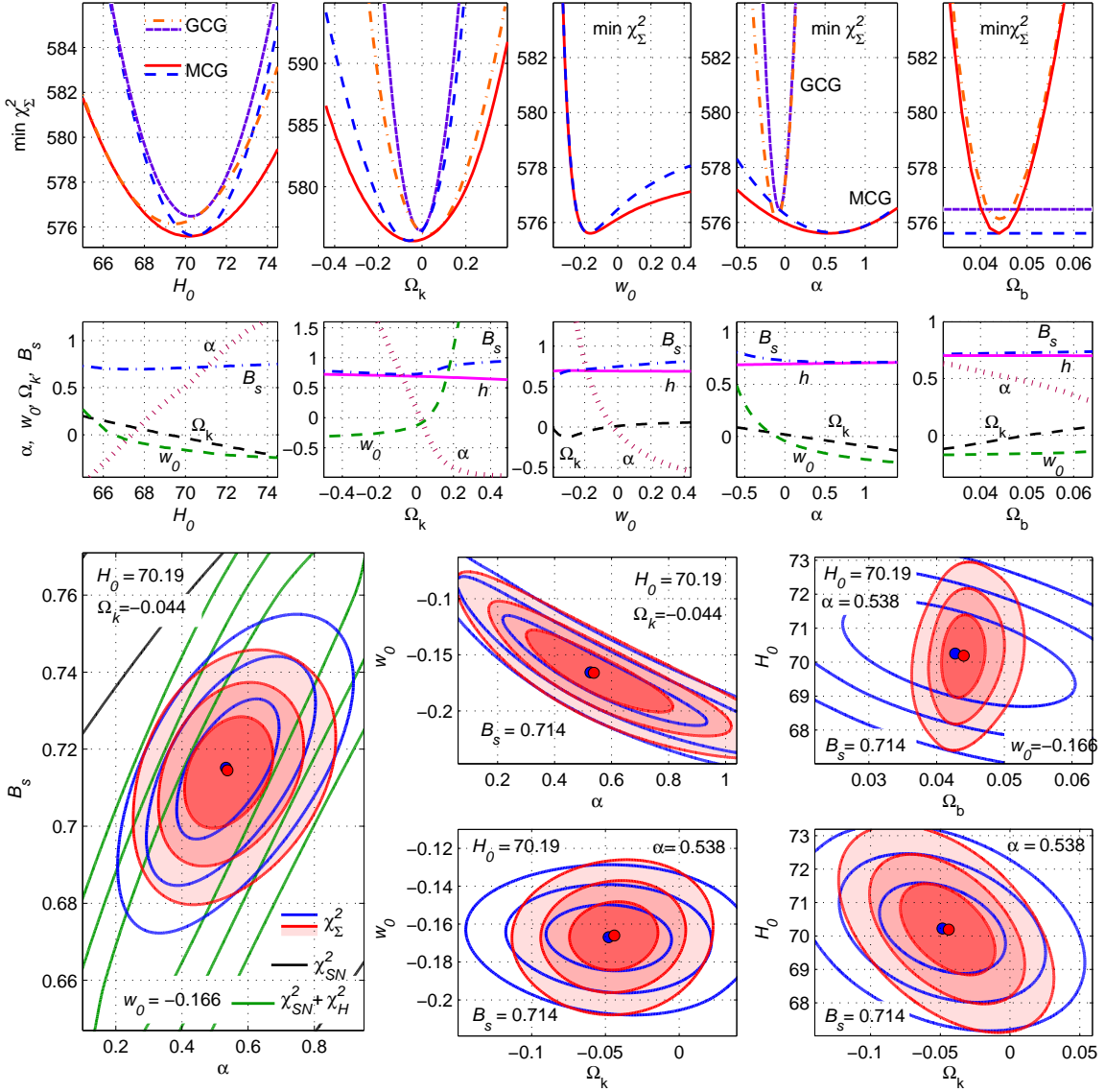}}
\caption{\small For the MCG model dependencies of $\min\chi^2_\Sigma$ on one chosen
parameter $H_0$, $\Omega_k$, $w_0$, $\alpha$, $\Omega_b$ are presented for the fitting
formulae (\ref{rsA}) (red solid lines) and (\ref{rsh}) (blue dashed lines) in comparison
with the GCG model (orange dash-dotted lines and violet dashed lines for these two
choices of $r_d$). In the second row there are correspondent MCG parameters of a minimum
point for $\chi^2_\Sigma$ with Eq.~(\ref{rsA}). In other panels for the MCG model
$1\sigma$, $2\sigma$, $3\sigma$ level lines are drawn in the planes of 2 parameters
(other parameters are fixed). Notations correspond to figure~\ref{F1}. }
  \label{F2}
\end{figure}

The MCG and GCG  models are to be explored as two-component models with  usual dust-like
baryonic matter component $\rho_b$ and the Chaplygin gas component $\rho_g$ with EoS
(\ref{EoSM}). In this case the total density (\ref{rho}) is
\begin{equation}
\rho=\rho_b+\rho_g,\qquad p_b=0.
 \label{rho2}
 \end{equation}
 However the first component $\rho_b$ and the corresponding fraction $\Omega_b$ (\ref{Omegab})
 may include not only visible baryonic matter but also a
part of cold dark matter with $\rho=\rho_{dm}$. Our practical applications of these
models in Ref.~\cite{ShV14} and in this paper (see the top-right panel of
figure~\ref{F2}) demonstrate rather weak dependence of $\min\chi^2_\Sigma$ on $\Omega_b$
for the model assumption (\ref{rsh}), but the strong $\Omega_b$ dependence for the
fitting formula (\ref{rsA}). This behavior resembles the $\Lambda$CDM model, where
separation of baryonic and cold dark matter in $\Omega_m$ appears only in
Eq.~(\ref{rsA}). In the MCG and GCG  models these matter fractions may also be mixed in
their observational manifestations, so below we did not use $\Omega_b$ as an usual free
parameter of the theory, but fix its fiducial value (\ref{Omb}) in the main part this
research (except for calculations, presented in right panels if figures~\ref{F2} and
\ref{F3}).

Equation (\ref{Esys2}) for the MCG model (\ref{EoSM}) is integrable, so the analog of
Eq.~(\ref{EqLCDM}) for this model is
\cite{Benaoum02,Chimento04,DebnathBCh04,LiuLiC05,LuXuWL11,PaulTh13,PaulTh14}
 \begin{equation}
\frac{H^2}{H_0^2}=
\Omega_b \Big(\frac{a}{a_0}\Big)^{\!-3}+
 \Omega_k\Big(\frac{a}{a_0}\Big)^{\!-2}+
 (1-\Omega_b-\Omega_k)\bigg[B_s+(1-B_s)\Big(\frac{a}{a_0}\Big)^{\!-3(1+w_0)(1+\alpha)}\bigg]^{1/(1+\alpha)}.
  \label{EqMCG}\end{equation}
  Here the dimensionless parameter $B_s=B\rho_0^{-1-\alpha}/(1+w_0)$ is used
instead of $B$. Thus in the MCG model (\ref{EoSM}) we have 6 independent parameters:
$H_0$, $\Omega_b$, $\Omega_k$, $w_0$, $\alpha$ and $B_s$. Naturally, the GCG model has 5
parameters: the same set without $w_0$.

In the MCG model the parameter $\Omega_m$ from its formal definition (\ref{Omega1})
equals $\Omega_m=1-\Omega_k$  in accordance with $\Omega_\Lambda=0$ and
Eq.~(\ref{sumOm}). However, the expression $A(z)$ (\ref{dzAz}) has the factor
$\sqrt{\Omega_m}$, so one should use the effective value $\Omega_m^{eff}$ in any model.
If we compare the early universe limit $z\gg1$ in the MCG equation (\ref{EqMCG}) with
the $\Lambda$CDM equation (\ref{EqLCDM}), we obtain the 
effective value \cite{LuXuWL11,PaulTh13,PaulTh14}:
\begin{equation}
 \Omega_m^{eff}=\Omega_b+(1-\Omega_b-\Omega_k)(1-B_s)^{1/(1+\alpha)}.
 \label{Ommeff1} \end{equation}

But for the majority of observational data in Ref.~\cite{SNTable} and tables~\ref{TBAO},
\ref{TH} redshifts are $0<z<1$, so to describe correctly these data we have to consider
the present time limit of Eq.~(\ref{EqMCG}). If we compare limits of the right hand
sides of Eqs.~(\ref{EqLCDM}) and (\ref{EqMCG}) at $z\to0$ or ${\cal A}\to0$, we obtain
another effective value \cite{ShV14}
\begin{equation}
 \Omega_m^{eff}=\Omega_b+(1-\Omega_b-\Omega_k)(1-B_s)(1+w_0).
 \label{Ommeff2} \end{equation}

Expressions (\ref{Ommeff1}) and (\ref{Ommeff2}) and their contributions in
$\chi^2_{BAO}$ and $\chi^2_\Sigma$ were compared in Ref.~\cite{ShV14}. In this paper we
use below Eq.~(\ref{Ommeff2}) and its analogs for other models.

Figure~\ref{F2} shows how  the MCG model (\ref{EoSM}) describes the observational data
from Ref.~\cite{SNTable} and tables~\ref{TBAO}, \ref{TH} in comparison with the GCG
model (\ref{EoSGCG}). Notations are the same as in figure~\ref{F1}. In the top line
panels  red solid lines and blue dashed lines denote graphs of $\min\chi^2_\Sigma$ for
the MCG model with expressions (\ref{rsA}) and (\ref{rsh}) correspondingly. The similar
functions for the GCG model are shown as orange dash-dotted lines, if $r_s(z_d)$ is
calculated with the formula (\ref{rsA}) and as violet dashed lines for the case
(\ref{rsh}).

These graphs in the top line describe how $\chi^2_\Sigma$ for the mentioned models
depends on one chosen model parameter: $H_0$, $\Omega_k$, $w_0$, $\alpha$, $\Omega_b$.
All these curves determine the optimal values and errors of the model parameters in
tables~\ref{Estim},~\ref{Estimrh}. The second row panels of figure~\ref{F2} correspond
to panels in the top line and present dependencies of coordinates of minima points on
$H_0,\dots$, $\Omega_b$ for $\chi^2_\Sigma$ in the MCG model with the fitting formula
(\ref{rsA}).

In the top-right panel of figure~\ref{F2} the value $\min\chi^2_\Sigma$ for the MCG
model means the minimum over other 5 parameters for each fixed $\Omega_b$:
$\min\chi^2_\Sigma=\min\limits_{H_0,\Omega_k,w_0,\alpha,B_s}\chi^2_\Sigma$. Our
calculations support the previous conclusion \cite{ShV14} about weak dependence of
$\min\chi^2_\Sigma$ on $\Omega_b$ for the case (\ref{rsh}) with $r_d=(r_d h)_{fid}\cdot
h^{-1}$ as for the the fixed value  $r_d$ (\ref{rs}) for both models. This is connected
with possible mixing of baryonic and cold dark matter in $\Omega_b$. However, for and
the fitting formula (\ref{rsA}) we see the sharp dependence of $\min\chi^2_\Sigma$ on
$\Omega_b$; for both  MCG and GCG models this picture coincides with  behavior of the
$\Lambda$CDM model in figure~\ref{F0}. As mentioned above, we will not use $\Omega_b$ as
an usual free model parameter and we fix its $\Lambda$CDM fiducial value (\ref{Omb}). It
is interesting that for the MCG and GCG models this value $\Omega_b\simeq0.044\pm0.004$
is very close and practically coincides with the $\Lambda$CDM fiducial value of this
parameter.

In figure~\ref{F2} the parameter $\Omega_b$ is varied only in 3 right panels, in other
12 panels this value is fixed in the form (\ref{Omb}). In particular, the value
$\min\chi^2_\Sigma(H_0)$ for the MCG model in the top-left panel is the minimum over 4
parameters
 $\min\limits_{\Omega_k,w_0,\alpha,B_s}\chi^2_\Sigma$ with fixed
$\Omega_b=0.044$. The similar picture takes place, when we study dependence of these
minimal functions on $\Omega_k$, $w_0$, $\alpha$, for example,
 $\min\chi^2_\Sigma(\Omega_k)=\min\limits_{H_0,w_0,\alpha,B_s}\chi^2_\Sigma$.
For the GCG model with $w_0=0$ these minima (orange dash-dotted lines and violet dashed
lines) are taken over 3 remaining parameters.

The graphs of $\min\chi^2_\Sigma(H_0)$ in the top-left panel with two expressions for
$r_s(z_d)$ resemble the $\Lambda$CDM case in figure~\ref{F1} for both MCG and GCG
models. For all three considered models this curve for the case  (\ref{rsh}) is slightly
more sharp  than for the formula (\ref{rsA}), this behavior may be seen in
tables~\ref{Estim}, \ref{Estimrh}, where all models are compared. Unlike the
$\Lambda$CDM model, for the MCG and GCG models we have different optimal values of $H_0$
(and also of $\Omega_k$, $\alpha$ and other parameters) in the cases  (\ref{rsA}) and
(\ref{rsh}).

Obviously that for the GCG model the graphs of $\chi^2_\Sigma$ in the top panels of
figure~\ref{F2} lie higher, then the correspondent MCG graphs. These curves converge at
points, where the optimal value of $w_0$ equals zero in the panel below.

The graphs of $\min\chi^2_\Sigma(\Omega_k)$  in the second top panel are asymmetric, for
$\Omega_k>0$ these values for the GCG and MCG models are close, but for  $\Omega_k<0$
the curves for these models diverge for both variants of $r_s(z_d)$. In the third top
panel we have asymmetric dependence of these minima on $w_0$ for the MCG model.

Dependence od $\min\chi^2_\Sigma$ on $\alpha$ is essentially different for the
considered models: for the GCG model these curves have sharp minima, in particular, for
the case (\ref{rsA})  at $\alpha\simeq-0.1$ with $\sigma\simeq0.1$, but for the MCG
model the dependence is rather smooth, the minimum at $\alpha\simeq0.54$ has
$\sigma\simeq0.9$. These results are also presented in table~\ref{Estim}, correspondent
values for the function (\ref{rsh}) are in table~\ref{Estimrh}. One may conclude that
changes of the parameter $w_0$ may compensate changes of $\alpha$.

In 5 bottom panels of figure~\ref{F2} for the MCG model we present $1\sigma$, $2\sigma$
and $3\sigma$ level lines in the planes of 2 model parameters (in notations of
figure~\ref{F1}), in particular, red filled contours denote levels of  $\chi^2_\Sigma$
with the fitting formula (\ref{rsA}). For this case in the bottom-left panel in the
$(\alpha,B_s)$ plane contours for  $\chi^2_{SN}$ (black) and $\chi^2_{SN}+\chi^2_H$
(green lines) are also shown. Contours for $\chi^2_\Sigma$ with the formula (\ref{rsh})
in all panels are drawn as blue lines. Other model parameters are fixed and shown in the
panels and in table~\ref{Estim}, they are optimal for the fitting formula (\ref{rsA})
(but not optimal for the case (\ref{rsh})).


 \begin{table}[ht]
 \centering
 {\begin{tabular}{||l||c||c|c|l||}  \hline
 Model   &$ \min\chi^2_\Sigma $&$H_0$&$\Omega_k$\,&other parameters  \\ \hline
 $ \Lambda$CDM$ $& 577.39& $70.07\pm1.82$ & $-0.028_{-0.048}^{+0.050}$ & $\Omega_m$=$\,0.278\pm0.008$,
$\Omega_\Lambda=0.750_{-0.055}^{+0.051}$\rule{0pt}{1.2em} \\ 
 \hline
 GCG & 576.13 & $69.46\pm1.88$ & $0.026_{-0.069}^{+0.075}$ & $\alpha= -0.100_{-0.098}^{+0.090},\; B_s=0.738_{-0.022}^{+0.024}$\rule{0pt}{1.2em}\\
   \hline 
 MCG & 575.60 & $70.19\pm2.15$ & $-0.044\pm0.122$ & $\alpha=0.538_{-0.893}^{+0.902},\;\; B_s=0.714_{-0.031}^{+0.033},$\rule{0pt}{1.2em}\\
    &   &   &  &$w_0=-0.166_{-0.089}^{+0.322}$\rule{0pt}{1.2em}  \\
  \hline
 EoS& 575.15 & $70.28_{-2.10}^{+2.22}$ & $-0.045_{-0.094}^{+0.102}$ & $p_0=-0.904_{-0.248}^{+0.256},\;\,\beta=-0.042_{-0.038}^{+0.041},
 $\rule{0pt}{1.2em} \\
  (\ref{EoS2}) &   &   &  &$w_0=0.183\pm0.222$  \\
 \hline \end{tabular}
 \caption{Models and $1\sigma$ estimates of model parameters, if $r_s(z_d)$ has the
form (\ref{rsA}); $\Omega_b=0.044$.}
 \label{Estim}}\end{table}

 \begin{table}[ht]
 \centering
 {\begin{tabular}{||l||c||c|c|l||}  \hline
  Model     &$ \min\chi^2_\Sigma $&$H_0$&$\Omega_k$&\,other parameters  \\ \hline
 $ \Lambda$CDM$ $& 577.39& $70.07\pm1.27$ & $-0.028_{-0.034}^{+0.035}$ & $\Omega_m$=$\,0.278\pm0.009$,
$\Omega_\Lambda=0.750_{-0.034}^{+0.033}$\rule{0pt}{1.2em} \\
\hline
 GCG & 576.48 & $70.27\pm1.28$ & $-0.009\pm0.040$ & $\alpha= -0.069\pm0.074,\;B_s=0.753\pm0.013$\\
   \hline
 MCG & 575.61 & $70.44\pm1.30$ & $-0.060_{-0.075}^{+0.068}$ & $\alpha=0.613_{-0.744}^{+0.80},\;\; B_s=0.716_{-0.033}^{+0.058},$\rule{0pt}{1.2em}  \\
    &   &  &  &$ w_0=-0.176_{-0.078}^{+0.447}$\rule{0pt}{1.2em}  \\
  \hline
 EoS & 575.14 & $70.42\pm1.30$ &$-0.048_{-0.048}^{+0.054}$&
 $p_0=-0.928_{-0.230}^{+0.242},\;\,\beta=-0.045_{-0.039}^{+0.043},$\rule{0pt}{1.2em}  \\
  (\ref{EoS2})  &   &  &  &$w_0=0.205_{-0.180}^{+0.187}$\rule{0pt}{1.2em}  \\
 \hline \end{tabular}
 \caption{Models and $1\sigma$ estimates of model parameters with  $r_s(z_d)$ from
 Eq.~(\ref{rsh});
$\Omega_b=0.044$.}
 \label{Estimrh}}\end{table}


For the GCG model (\ref{EoSGCG}) the values in table~\ref{Estimrh} are close to our
previous estimations \cite{ShV14} $H_0=70.093\pm0.369$, $\Omega_k=-0.019\pm0.045$,
$\alpha= -0.066_{-0.074}^{+0.072}$, $B_s=0.759_{-0.016}^{+0.015}$ with 7 and 34 data
points for $d_z(z)$ and $H(z)$ correspondingly.

These estimates for the MCG model should be compared with similar results for this model
in papers \cite{LuXuWL11,PaulTh13,PaulTh14}. The authors of Ref.~\cite{LuXuWL11} for the
flat model with $\Omega_k=0$ described the observational data with 557, 15 and 2 data
points correspondingly for supernovae, $H(z)$ and BAO, but they also included the
cluster X-ray gas mass fraction data. Their $1\sigma$ estimations
$H_0=70.711_{-3.142}^{+4.188}$ and $B_s=0.7788_{-0.0723}^{+0.0736}$ are more wide then
ours in tables~\ref{Estim}, \ref{Estimrh}; however $\alpha=0.1079_{-0.2539}^{+0.3397}$
and the narrow box $w_0=0.00189_{-0.00756}^{+0.00583}$ lie inside our $1\sigma$
estimates.

In Refs.~\cite{PaulTh13,PaulTh14} the MCG model with $\Omega_k=0$ is applied for
describing 12 $H(z)$ data points, 11 points for the growth function $f=d\log\delta/d\log
a$ of the large scale structures, 17 points for $\sigma_8(z)$ and also in
Ref.~\cite{PaulTh14} observations of supernovae and 1 data point for BAO. The authors
did not demonstrate their estimates for $H_0$, but noted ambiguously ``$\chi^2$ function
for the background test is minimized
 by the present Hubble value predicted by WMAP7''. The best fit values of other
parameters in  Ref.~\cite{PaulTh14} are $w_0=0.005$, $\alpha=0.19$, $B_s=0.825$ with
errors, calculated for pairs of these parameters. Only the estimate for $B_s$ is in
tension with our results in tables~\ref{Estim}, \ref{Estimrh}.

\section{Model with quadratic equation of state}\label{Quad}

It is interesting to compare  the MCG model and the model with quadratic equation of
state \cite{NojOdin04,NojOdinTs05,AnandaB06,LinderS09,AstashNojOdinY12,Chavanis13}
 \begin{equation}
 p_g=\tilde p_0+w_0\rho_g+\tilde\beta\rho_g^2,
 \label{EoS20}
  \end{equation}
because both models have 6 parameters: $H_0$, $\Omega_k$, $\Omega_b$ and 3 parameters in
the EoS (\ref{EoSM}) or (\ref{EoS20}). It is convenient to use the critical density
$\rho_{cr}=3H_0^2/(8\pi G)$, introduce the dimensionless parameters $p_0=\tilde
p_0/\rho_{cr}$, $\beta=\tilde\beta\rho_{cr}$
 instead of $\tilde p_0$, $\tilde\beta$ and rewrite  Eq.~(\ref{EoS20})
in the form
 \begin{equation}
 p_g=p_0\rho_{cr}+w_0\rho_g+\beta\rho_g^2/\rho_{cr}.
 \label{EoS2}
  \end{equation}

Similarly to the GCG and MCG models the model with quadratic EoS (\ref{EoS20}) or
(\ref{EoS2}) has the analytical general solution of Eq.~(\ref{Esys2}) \cite{AnandaB06}
 \begin{equation}
 \frac{\rho_g}{\rho_{cr}}=\left\{\begin{array}{ll}
 \frac1{2\beta}\bigg[\frac{\Gamma-\sqrt{|\Delta|}\tan\big(\frac32\sqrt{|\Delta|}\,{\cal A}\big)}
 {1+|\Delta|^{-1/2}\tan\big(\frac32\sqrt{|\Delta|}\,{\cal A}\big)}-1-w_0\bigg],& \Delta<0,\\
  \frac1{2\beta}\Big[\big(\frac32{\cal A}+1/\Gamma\big)^{-1}-1-w_0\Big],&  \Delta=0,\\
 \frac{\rho_-(\Omega_m-\rho_+)(a/a_0)^{-3\sqrt{\Delta}}-\rho_+(\Omega_m-\rho_-)}
 {(\Omega_m-\rho_+)(a/a_0)^{-3\sqrt{\Delta}}-\Omega_m+\rho_-},&
 \Delta>0,
 \end{array}\right.
 \label{rhopr} \end{equation}
 depending on a sign of
the discriminant $\Delta=(1+w_0)^2-4\beta p_0$. Here
 $$
\Omega_m=1-\Omega_k-\Omega_b,\quad\Gamma=2\beta\Omega_m+1+w_0,\quad
 \rho_\pm=\frac{-1-w_0\pm\sqrt{\Delta}}{2\beta},\quad {\cal A}=\log\frac a{a_0}.
 $$

 The equation (\ref{Esysa}) for this model is reduced to the  form
 \begin{equation}
 \frac{H^2}{H_0^2}=\bigg(\frac{d{\cal A}}{d\tau}\bigg)^2=
\frac{\rho_g}{\rho_{cr}}+ \Omega_b e^{-3{\cal A}}+\Omega_ke^{-2{\cal
 A}}.
 \label{Eq2} \end{equation}
We solve this equation numerically from the present time initial condition  ${\cal
A}\big|_{\tau=1}=0$ ``to the past''. We can use analytical solution (\ref{rhopr}) or
solve Eq.~(\ref{Esys2}) numerically, these approaches are equivalent.

For the model (\ref{EoS2}) the effective value
 \begin{equation}
 \Omega_m^{eff}=\Omega_b+p_0+\Omega_m(1+w_0+\beta\Omega_m)
 \label{Ommeffqu} \end{equation}
is calculated in the $z\to0$ or ${\cal A}\to0$ limit similarly to the MCG model.

The model with quadratic EoS (\ref{EoS2}) in the domain $\beta>0$ may have  the
following singularity in the past: when $t\to t_*+0$, density grows to infinity, but the
scale factor remains finite and nonzero: $\lim\limits_{t\to t_*}\rho_g=\infty$,
$\lim\limits_{t\to t_*}a=a(t_*)\ne0$. This behavior resembles the Type III finite-time
future singularity from the classification \cite{Bamba12,NojOdinFR}. In
Ref.~\cite{Chavanis13} the author did not see these singularities, because he considered
only negative values $\beta=-(w_0+1)/\rho_P$.

\begin{figure}[th]
  \centerline{\includegraphics[scale=0.72,trim=5mm 0mm 5mm -1mm]{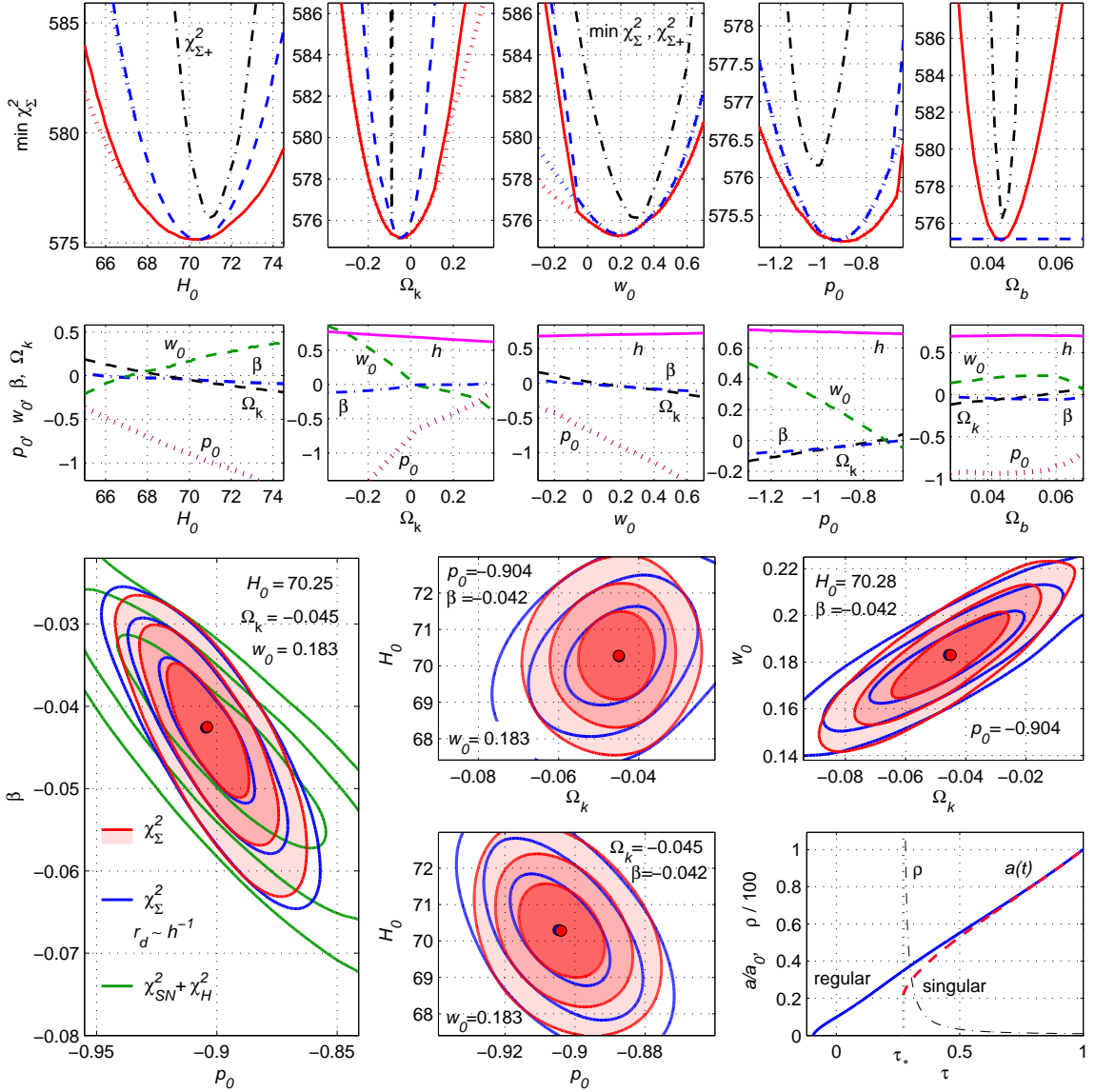}}
  \caption{\small For the model with quadratic EoS (\ref{EoS2}) one parameter dependencies
of $\min\chi^2_\Sigma$ with $r_s(z_d)$ in the forms (\ref{rsA}) (red solid lines) and
(\ref{rsh}) (blue dashed lines), of $\min\chi^2_\Sigma$ with CMB contribution and
Eq.~(\ref{rsA}) (black dash-dotted lines) and also coordinates of minima points, level
lines are presented in notations of figure~\ref{F2}. In the bottom-right panel scale
factors $a/a(0)$ for the regular solution (the blue solid line) and the singular
solution (the red dashed line) are shown.}
  \label{F3}
\end{figure}

In the bottom-right panel of figure~\ref{F3} we present the example of singular solution
as the red dashed line for $a(\tau)/a_0$ and the black dash-dotted line for
$0.01\cdot\rho(\tau)/\rho_{cr}$. This singularity is compared with the regular solution
(the blue solid line for $a(\tau)/a_0$) with the optimal values of model parameters from
table~\ref{Estim}. For the singular solution in this panel $\beta=0.02$, but other model
parameters are from table~\ref{Estim}.

We have to exclude these nonphysical singular solutions, for this purpose we can use
different approaches. The simplest way is to add the penalty contribution in
$\chi^2_\Sigma$ in the form
 \begin{equation}
 \Delta\chi^2_\Sigma=P_1\big[\exp\big(P_2a(t_*)/a_0\big)-1\big].
  \label{penal} \end{equation}
The function (\ref{penal}) with $P_1=0.3$ and $P_2=20$ successfully helps to avoid this
singularity. For regular solutions $a(t_*)=0$ and the contribution (\ref{penal})
vanishes. In the top panels of  figure~\ref{F3} functions $\min\chi^2_\Sigma$ with the
contribution (\ref{penal}) are shown as solid red lines for the formula (\ref{rsA}) and
as blue dashed lines for $r_d=(r_d h)_{fid}\cdot h^{-1}$; for the case without penalty
(\ref{penal}) these graphs are dotted lines of the correspondent color: red for the
formula (\ref{rsA}) and blue for (\ref{rsh}).

More natural method to eliminate nonphysical singular solutions is to include early time
parameters (for example, $z_d$) into consideration. Unfortunately, both equations
(\ref{rsA}) and (\ref{rsh}) or $r_s(z_d)$ are insensitive to the mentioned
singularities. So we can take into account the  cosmic microwave background (CMB)
constraints,  in particular, for the values \cite{Aubourg14}
 \begin{equation}
  \mathbf{v}=\big(\omega_b,\;\omega_{cb},\;D_M(1090)/r_d\big),
  \label{CMB} \end{equation}
where $\omega_i=\omega_ih^2$, $D_M(z)=D_L(z)\big/(1+z)$. If we calculate the vector
(\ref{CMB}), compare it with the estimation \cite{Aubourg14,WMAP}
 $\Delta \mathbf{v}=\mathbf{v}-\big( 0.02259,\; 0.1354,\;94.51\big)$
and add the corresponding term to the $\chi^2_\Sigma$ function (\ref{chisum}), we obtain
  \begin{equation}
\chi^2_{\Sigma+}=\chi^2_\Sigma+\Delta\mathbf{v}\cdot
C_{CMB}^{-1}\big(\Delta\mathbf{v}\big)^{T}.
 \label{chiCMB} \end{equation}
Here the covariance matrix from Ref.~ \cite{Aubourg14} contains
$c_{11}=2.864\cdot10^{-7}$, $c_{12}=-4.809\cdot10^{-7}$, $c_{13}=-1.111\cdot10^{-5}$,
 $c_{22}=1.908\cdot10^{-5}$,  $c_{23}=-7.495\cdot10^{-6}$,  $c_{33}=0.02542$.

Graphs of one parameter functions $\min\chi^2_{\Sigma+}$ with the CMB contribution for
the case (\ref{rsA}) are also presented in the top panels of  figure~\ref{F3} as black
dash-dotted lines.

We demonstrate in figure~\ref{F3} how the model (\ref{EoS2}) is effective in describing
the observational data from Ref.~\cite{SNTable} and tables~\ref{TBAO}, \ref{TH}.
Notations are the same as in figure~\ref{F2}. Here we also fix the value (\ref{Omb})
$\Omega_b=0.044$ (except for 2 top-right panels) and do not use $\Omega_b$ as a fitting
parameter. The dependence
$\min\chi^2_\Sigma(\Omega_b)=\min\limits_{H_0,\Omega_k,w_0,p_0,\beta}\chi^2_\Sigma$
(shown in the top-right panel) is rather weak for the case (\ref{rsh}) $r_d=(r_d
h)_{fid}\cdot h^{-1}$, but it is essential for the formula (\ref{rsA}) and for the
function $\min\chi^2_{\Sigma+}$. In the  case $\min\chi^2_\Sigma$ with Eq.~(\ref{rsA})
(the red curve) the dependence $\min\chi^2_\Sigma(\Omega_b)$ is very close to the
correspondents functions for the $\Lambda$CDM, GCG, MCG models in figures~\ref{F0},
\ref{F2}; its minimum is also at $\Omega_b\simeq0.044\pm0.004$.

In the top line panels of figure~\ref{F3} we draw graphs of $\chi^2_\Sigma$ minima
depending on one model parameter ($H_0$, $\Omega_k\dots$) for the model (\ref{EoS2})
with the penalty function  (\ref{penal}) as solid red lines for the formula (\ref{rsA})
and as blue dashed lines for the case (\ref{rsh}). Correspondent lines  without
contribution (\ref{penal}) are shown as dots of the same color.  In other words, if we
calculate $\min\chi^2_\Sigma$  only for (physical) regular solutions, we obtain the
solid or dashed lines; for dots we also include singular solutions without physical
interpretation. These lines coincide in domains where best values of $\beta$ are
negative and corresponding solutions are regular. One can see in figure~\ref{F3} that
optimal values of model parameters correspond to regular solutions in both considered
cases (\ref{rsA}) (table~\ref{Estim}) and (\ref{rsh}) (table~\ref{Estimrh}).

Minima of $\chi^2_\Sigma$ and  $\chi^2_{\Sigma+}$ here have the same sense as in
figure~\ref{F2} for the MCG model, in particular, in the top-left panel
$\min\chi^2_\Sigma(H_0)=\min\limits_{\Omega_k,w_0,p_0,\beta}\chi^2_\Sigma$. Dependence
of this minimum on $H_0$ for two formulas for  $r_s(z_d)$ resembles other considered
models (compare with figures~\ref{F1},~\ref{F2}): the graph for the case (\ref{rsh}) is
more sharp, than for Eq.~(\ref{rsA}), it corresponds to larger optimal value of $H_0$
and smaller $1\sigma$ error in table~\ref{Estimrh} than the values in table~\ref{Estim}.
For other panels of the top line in figure~\ref{F3} dependence of $\min\chi^2_\Sigma$ on
parameters $\Omega_k$, $w_0$ and $p_0$ results in the correspondent the $1\sigma$
estimates in tables~\ref{Estim},~\ref{Estimrh}.

The functions $\min\chi^2_{\Sigma+}$ with  the CMB contribution (\ref{chiCMB}) depending
on one parameter are shown as black dash-dotted lines in the top line panels. We see
that the absolute minimum of this function is $\min\chi^2_{\Sigma+}\simeq 576.16$, in a
bit exceeds the corresponding value $\min\chi^2_\Sigma\simeq 575.15$ for the case
(\ref{rsA}) in table~\ref{Estim}. However,  the CMB contribution in the form
(\ref{chiCMB}) works as a very narrow filter for some model parameters, in particular,
for the considered model with EoS (\ref{EoS2}) this contribution rigidly constrains the
value $\Omega_k$, so we have the $\chi^2_{\Sigma+}$ estimation
$\Omega_k=-0.090\pm0.001$. The curve, corresponding to this narrow range, is shown in
the second top panel of figure~\ref{F3}, it is too narrow, so it looks like a vertical
black segment.

The $\chi^2_{\Sigma+}$ estimations of other parameters are also more narrow (see the top
line in  figure~\ref{F3}), than for both cases of  $\chi^2_\Sigma$, it is connected with
restrictions for $\Omega_k$ and other correspondent parameters for other models. Some of
these restrictions look like artificial od connected with the concrete choice of the CMB
vector (\ref{CMB}). So further we consider the functions $\min\chi^2_{\Sigma}$ without
the CMB contribution (\ref{chiCMB}) with the fitting formula (\ref{rsA}) as the most
reliable indicator.

Panels in the second row of figure~\ref{F3} correspond the upper panels for the function
$\min\chi^2_\Sigma$ with the fitting formula (\ref{rsA})  with the penalty contribution
(\ref{penal}). They demonstrate evolution of coordinates of the minimal point if we vary
the chosen parameter. In 4 bottom panels functions $\chi^2$ depend on two model
parameters ($p_0,\beta$; $p_0,H_0$; $\Omega_k,H_0$; $\Omega_k,w_0$) when other
parameters are fixed with their optimal values from table~\ref{Estim}. Red filled
contours denote the case (\ref{rsA}),  blue lines corresponds to (\ref{rsh}). These
level lines demonstrate similarity with the MCG model in figure~\ref{F2}, but for the
model (\ref{EoS2}) we have alternative behavior in the ($\Omega_k,H_0$) plane. One
should emphasize that all $1\sigma$, $2\sigma$ and $3\sigma$ level lines for
$\chi^2_\Sigma$ in figure~\ref{F3} and optimal values of the model parameters in
table~\ref{Estim},~\ref{Estimrh} lie in the domain with regular solutions of the
quadratic EoS  model (\ref{EoS2}).

\section{Conclusion}

In this paper the Type Ia supernovae observational data from Ref.~\cite{SNTable} and
estimations of BAO parameters and $H(z)$ from tables~\ref{TBAO}, \ref{TH} are described
with the models $\Lambda$CDM, GCG (\ref{EoSGCG}), MCG (\ref{EoSM}) and the model with
quadratic EoS (\ref{EoS2}). Two approaches in estimation of the sound horizon scale
$r_s(z_d)$ are used and compared: the fitting formula (\ref{rsA}) with results,
tabulated in table~\ref{Estim}, and the expression (\ref{rsh}) $r_d=(r_d h)_{fid}\cdot
h^{-1}$ (table~\ref{Estimrh}.  Optimal values of model parameters with $1\sigma$ errors
in these tables are calculated via one-parameter distributions
(figures~\ref{F1}\,--\,\ref{F3}).

We also considered the CMB contribution in the form \cite{Aubourg14} (\ref{chiCMB}), the
results for  the model with EoS (\ref{EoS2}) are shown in figure~\ref{F3}. However, this
contribution appeared to be too sensitive and restrictive for some model parameters, in
particular,  $\Omega_k$ (or  $\Omega_\Lambda$ for the $\Lambda$CDM model). So we
consider the $\chi^2_{\Sigma}$ estimations of model parameters   in tables~\ref{Estim}
and \ref{Estimrh} as more reliable.

It is interesting that  predictions  of  the $\Lambda$CDM, GCG, MCG models and the model
with quadratic EoS for $\Omega_b$ are very close ($\Omega_b=0.044\pm0.004$),  if we
adopt  the fitting formula (\ref{rsA}) for $r_s(z_d)$. We use this fact and do not
consider $\Omega_b$ as an usual model parameter and fix its value in the form
(\ref{Omb}). One should note, that  predictions  of different models for $H_0$ and
$\Omega_k$ in tables~\ref{Estim} and \ref{Estimrh} are also rather close.

Absolute minima of $\chi^2_\Sigma$ with  the formula (\ref{rsA}) in table~\ref{Estim}
differ from the correspondent  minima of in table~\ref{Estimrh} with Eq.~(\ref{rsh}),
but the hierarchy of all considered models is the same in these tables. In particular,
the absolute minimum of $\chi^2_\Sigma$ in table~\ref{Estim} vary from the worst value
$577.39$ for the $\Lambda$CDM to  the best result $575.15$ for the model with quadratic
EoS (\ref{EoS2}). Note that the advantage of the MCG model in comparison with GCG is
larger in the case (\ref{rsh}) in table~\ref{Estimrh}.

  However, effectiveness of a model essentially depends on its number $N_p$
 of model parameters (degrees of freedom). This number is used in model selection
statistics, in particular, in the following Akaike information criterion
\cite{ShiHL12,Szydlten06}
 $$
 AIC = \min\chi^2_\Sigma +2N_p.
 $$

If we fix the value $\Omega_b$ in the form (\ref{Omb}) for the models GCG, MCG and with
EoS (\ref{EoS2}) and do not use this parameter as a degree of freedom, we will have the
numbers  $N_p$ and $AIC$ for  $\chi^2_\Sigma$ from table~\ref{Estim} for the considered
models tabulated here in table~\ref{AIC}.
\begin{table}[h]
\centering
 {\begin{tabular}{||l||c|c|c||}  \hline
  Model       &$\min\chi^2_\Sigma$ &$N_p$&$AIC$ \\ \hline
 $\Lambda$CDM & 577.39& 3 &583.39 \\ \hline
 GCG & 576.13 & 4 &  584.13   \\
  \hline
 MCG & 575.60 & 5 & 585.60 \\
  \hline
 EoS (\ref{EoS2})& 575.15 & 5 & 585.15  \\
 \hline \end{tabular}
 \caption{Akaike information criterion for the models.}
 \label{AIC}}\end{table}

This information criterion works against models with large $N_p$ and adds arguments in
favor of the $\Lambda$CDM model.

\section*{Acknowledgments}

The work is supported by the Ministry of education and science of Russia, grant No.
1686. The author is grateful to S. D. Odintsov and E. G.  Vorontsova for useful
discussions, and to two unknown referees for valuable advices.


\begin{thebibliography}{22}

\bibitem{Riess98}  A. G. Riess { et al.}, 
{\it Astron. J.} {\bf 116} (1998) 1009, arXiv:astro-ph/9805201.

\bibitem{Perl99} S. Perlmutter { et al.}, 
{\it Astrophys. J.} {\bf517} (1999) 565,
 arXiv:astro-ph/9812133.

\bibitem{SNTable}
N. Suzuki { et al.}, 
{\it Astrophys. J.} {\bf746} (2012) 85, arXiv:1105.3470 [astro-ph.CO];
http://supernova.lbl.gov/Union/.

\bibitem{WeinbergAcc12}
D. H. Weinberg { et al.}, 
{\it Phys. Rep.} {\bf530}, (2013) 87, arXiv: 1201.2434 [astro-ph.CO].

\bibitem{Eisen05}
D. J. Eisenstein { et al.}, 
{\it Astrophys. J.} {\bf633} (2005) 560, arXiv:astro-ph/0501171.

\bibitem{EisenHu98}
D. J. Eisenstein and W. Hu, 
 {\it Astrophys. J.} {\bf496} (1998)  605, arXiv:astro-ph/9709112.

\bibitem{Aubourg14}
E. Aubourg { et al.}, 
{\it Phys. Rev. D} {\bf 92} (2015) 123516, arXiv:1411.1074 [astro-ph.CO].

\bibitem{WMAP} WMAP collaboration, G. Hinshaw { et al.},
 {\it Astrophys. J. Suppl.} {\bf 208} (2013)  19, arXiv:1212.5226 [astro-ph.CO].

\bibitem{Plank13}
Planck Collaboration, P. A. R. Ade { et al.} 
 {\it Astron. Astrophys.}  {\bf571} (2014) A16, arXiv:1303.5076 [astro-ph.CO].

\bibitem{Plank15}
Planck Collaboration, P. A. R. Ade { et al.} 
 arXiv:1502.01589 [astro-ph.CO].

\bibitem{Simon05}
 J. Simon, L. Verde and R. Jimenez, 
{\it Phys. Rev.  D} {\bf 71} (2005) 123001, arXiv:astro-ph/0412269.

\bibitem{Stern10}
D. Stern,  R. Jimenez,  L. Verde, M. Kamionkowski and S. A. Stanford,
 {\it  J. Cosmol. Astropart. Phys.} {\bf 02} (2010) 008, arXiv:0907.3149 [astro-ph.CO].

\bibitem{Moresco12}
M. Moresco { et al.}, 
{\it J. Cosmol. Astropart. Phys.} {\bf 8} (2012) 006, arXiv:1201.3609
 [astro-ph.CO].

\bibitem{Zhang12}
 C. Zhang { et al.}, 
{\it Res. Astron. Astrophys.} {\bf 14} (2014) 1221, 
 arXiv:1207.4541 [astro-ph.CO].

\bibitem{Moresco15}
M. Moresco, 
 arXiv:1503.01116 [astro-ph.CO].

\bibitem{Gazta09}
E. Gazta\~naga,  A. Cabre, L. Hui, 
 {\it Mon. Not. Roy. Astron. Soc.} {\bf 399(3)} (2009) 1663. 
arXiv:0807.3551 [astro-ph].

\bibitem{Blake12}
C. Blake { et al.}, 
{\it Mon. Not. Roy. Astron. Soc.} {\bf 425(1)} (2012) 405, 
arXiv:1204.3674 [astro-ph.CO].

\bibitem{Busca12}
 N. G. Busca { et al.}, 
 {\it Astron. and Astrop.} {\bf 552} (2013) A96 arXiv:1211.2616 [astro-ph.CO].

\bibitem{ChuangW12}
C-H. Chuang and Y. Wang, 
 {\it Mon. Not. Roy. Astron. Soc.} {\bf 435(1)} (2013) 255, 
arXiv:1209.0210 [astro-ph.CO].

 \bibitem{Chuang13}
C-H. Chuang { et al.}, 
 {\it Mon. Not. Roy. Astron. Soc.} {\bf 433(4)} (2013) 3559, 
arXiv:1303.4486 [astro-ph.CO].

\bibitem{Anderson13}
L. Anderson { et al.}, 
 {\it Mon. Not. Roy. Astron. Soc.} {\bf 439(1)} (2014) 83, 
 arXiv:1303.4666 [astro-ph.CO].

\bibitem{Anderson14}
L. Anderson { et al.},  
 {\it Mon. Not. Roy. Astron. Soc.} {\bf 441} (2014) 24,
  arXiv:1312.4877 [astro-ph.CO].

\bibitem{Oka13}
A. Oka { et al.}, 
 {\it Mon. Not. Roy. Astron. Soc.} {\bf 439(3)} (2014) 2515, 
arXiv:1310.2820 [astro-ph.CO].

\bibitem{Font-Ribera13}
A. Font-Ribera { et al.}, 
 {\it J. Cosmol. Astropart. Phys.} {\bf 05} (2014) 027, arXiv:1311.1767 [astro-ph.CO].

\bibitem{Delubac14}
T. Delubac { et al.}, 
 {\it Astron. and Astrop.} {\bf 574} (2015) A59, 
arXiv:1404.1801 [astro-ph.CO].

\bibitem{Percival09}
 W. J. Percival { et al.},
{\it Mon. Not. Roy. Astron. Soc.} \textbf{401} (2010) 2148, 
arXiv:0907.1660 [astro-ph.CO].

\bibitem{Kazin09}
E. A. Kazin { et al.}, 
{\it  Astrophys. J.} {\bf710} (2010) 1444, 
arXiv:0908.2598 [astro-ph.CO].

\bibitem{Beutler11}
F. Beutler { et al.}, 
 {\it Mon. Not. Roy. Astron. Soc.} {\bf 416} (2011) 3017,
arXiv:1106.3366 [astro-ph.CO].

\bibitem{BlakeBAO11}
 C. Blake { et al.}, 
 {\it Mon. Not. Roy. Astron. Soc.}  {\bf 418} (2011) 1707, 
arXiv:1108.2635 [astro-ph.CO].

\bibitem{Padmanabhan12}
N. Padmanabhan { et al.}, 
 {\it Mon. Not. Roy. Astron. Soc.}  {\bf 427} (2012) 2132,
arXiv:1202.0090 [astro-ph.CO].

\bibitem{Seo12}
H.-J. Seo  { et al.}, 
{\it Astrophys. J.} {\bf761} (2012) 13, arXiv:1201.2172 [astro-ph.CO].

\bibitem{Kazin14}
E. A. Kazin { et al.}, 
{\it Mon. Not. Roy. Astron. Soc.} {\bf 441} (2014) 3524, arXiv:1401.0358 [astro-ph.CO].

\bibitem{Ross14}
 A. J. Ross { et al.}, 
{\it Mon. Not. Roy. Astron. Soc.} {\bf 449} (2015) 835,
 arXiv:1409.3242 [astro-ph.CO].



\bibitem{CopelandST06} E. J. Copeland, M. Sami and S. Tsujikawa, 
{\it Int. J. Mod. Phys. D} {\bf15} (2006) 1753, hep-th/0603057.

\bibitem{Clifton}  T. Clifton, P. G. Ferreira, A. Padilla and C. Skordis,
{\it Physics Reports} {\bf 513} (2012) 1, arXiv:1106.2476 [astro-ph.CO].

\bibitem{Bamba12} K. Bamba, S. Capozziello, S. Nojiri and S.~D.~Odintsov,
{\it Astrophys. and Space Science} {\bf342} (2012) 155, 
 arXiv:1205.3421 [gr-qc].

\bibitem{NojOdinFR}
S. Nojiri and S. D. Odintsov,
{\it Phys. Rept.} {\bf505} (2011) 59, 
arXiv:1011.0544  [gr-qc].

\bibitem{GrSh13}
O. A. Grigorieva and G. S. Sharov,  
{\it Int. J. Mod. Phys. D} {\bf 22} (2013) 1350075, arXiv:1211.4992, [gr-qc].

\bibitem{ShV14}
G. S. Sharov and E. G. Vorontsova,
{\it J. Cosmol. Astropart. Phys.} {\bf 10} (2014) 057, arXiv:1407.5405, [gr-qc].

\bibitem{KamenMP01}
A. Y. Kamenshchik, U. Moschella and V. Pasquier,
{\it Phys. Lett. B} {\bf 511(2-4)} (2001) 265, 
arXiv:gr-qc/0103004.
\bibitem{Benaoum02}
H. B. Benaoum, 
arXiv:hep-th/0205140.

\bibitem{Chimento04}
L. P. Chimento, {\it Phys. Rev. D.} {\bf 69} (2004) 123517.

\bibitem{DebnathBCh04}
U. Debnath, A. Banerjee, S. Chakraborty,
{\it Class. Quant. Grav.} {\bf 21} (2004) 5609, arXiv:gr-qc/0411015. 

\bibitem{LiuLiC05}
D-J. Liu, X-Z. Li, 
{\it Chin. Phys. Lett.} {\bf 22} (2005) 1600, astro-ph/0501115.

\bibitem{LuXuWL11}
J. Lu, L. Xu, Y. Wu and M. Liu,
{\it Gen. Rel. Grav.} {\bf 43} (2011) 819, arXiv:1105.1870 [astro-ph.CO].

\bibitem{PaulTh13}
B. C. Paul and P. Thakur,
{\it J. Cosmol. Astropart. Phys.} {\bf 11} (2013) 052, arXiv:1306.4808 [astro-ph.CO].

\bibitem{PaulTh14}
B.C. Paul, P. Thakur  and A. Beesham,
arXiv:1410.6588  [astro-ph.CO].

\bibitem{NojOdin04}
S. Nojiri and S. D. Odintsov,
{\it Phys. Rev. D} {\bf70} (2004) 103522, arXiv:hep-th/0408170.

\bibitem{NojOdinTs05}
S. Nojiri, S. D. Odintsov and S. Tsujikawa,
{\it Phys. Rev. D} {\bf71} (2005) 063004, arXiv:hep-th/0501025.


\bibitem{AnandaB06}
K. N. Ananda and M. Bruni, 
{\it Phys. Rev. D} {\bf 74} (2006) 023523,  arXiv:astro-ph/0512224.

\bibitem{LinderS09}
 E. V. Linder and R. J. Scherrer,
 {\it Phys. Rev. D} {\bf 80} (2009) 023008, arXiv:0811.2797 [astro-ph].

\bibitem{AstashNojOdinY12}
A. V. Astashenok, S. Nojiri, S. D. Odintsov and A. V. Yurov,
{\it Phys. Lett. B} {\bf709} (2012) 396, arXiv:1201.4056 [gr-qc].

\bibitem{Chavanis13}
P-H. Chavanis, 
arXiv:1309.5784 [astro-ph.CO].

\bibitem{NessPeri05}
S. Nesseris and L. Perivolaropoulos, {\it Phys. Rev. D} {\bf 72} (2005) 123519,
arXiv:astro-ph/0511040.

\bibitem{FarooqMR13}
O. Farooq, D. Mania and B. Ratra, 
{\it Astrophys. J.} {\bf 764} (2013) 139,  arXiv:1211.4253 [astro-ph.CO].

\bibitem{FarooqR13}
O. Farooq and B. Ratra, 
{\it Astrophys. J.} {\bf 766} (2013) L7, arXiv:1301.5243 [astro-ph.CO].

\bibitem{ShiHL12}
 K. Shi, Y. F. Huang and T. Lu,
{\it Monthly Notices Roy. Astron. Soc.}  {\bf 426} (2012) 2452, arXiv:1207.5875
[astro-ph.CO].

\bibitem{Riess11}
A. G. Riess { et al.}, 
{\it Astrophys. J.} {\bf 730(2)} (2011) 119, arXiv:1103.2976 [astro-ph.CO].

\bibitem{Bento02}
M. C. Bento, O. Bertolami and  A. A. Sen, 
{\it Phys. Rev. D} {\bf 66(4)} (2002) 043507, arXiv:gr-qc/0202064.

\bibitem{Szydlten06}
M. Szyd\l owski, A. Kurek and A. Krawiec, 
{\it Phys. Lett. B} {\bf642} (2006) 171, arXiv:astro-ph/0604327.

\end{thebibliography}
\end{document}